\documentclass[twocolumn]{aastex7}

\usepackage{float}
\usepackage{amsmath}
\usepackage{soul}
\usepackage{hyperref}

\newcommand{\appropto}{\mathrel{\vcenter{
  \offinterlineskip\halign{\hfil$##$\cr
    \propto\cr\noalign{\kern2pt}\sim\cr\noalign{\kern-2pt}}}}}

\newcommand{\gtrpropto}{\mathrel{\vcenter{
  \offinterlineskip\halign{\hfil$##$\cr
    \propto\cr\noalign{\kern2pt}>\cr\noalign{\kern-2pt}}}}}

\newcommand{\msun}{\mathrm{M_{\odot}}} 
\newcommand{\msunyr}{\msun\,{\rm yr}^{-1}} 

\newcommand{\mhe}{M_{\mathrm{He}}} 
\newcommand{\mwd}{M_{\mathrm{WD}}} 

\newcommand{\mdot}{\dot{M}} 
\newcommand{\mdotdec}{\dot{M}_{\rm dec}}
\newcommand{\mdotw}{\dot{M}_{\rm w}}

\newcommand{\Lhe}{L_{\mathrm{He}}} 
\newcommand{\Lwd}{L_{\mathrm{WD}}} 

\newcommand{\Lacc}{L_{\mathrm{acc}}}

\newcommand{\Teffwd}{T_{\mathrm{eff,WD}}}
\newcommand{\Teff}{T_{\mathrm{eff}}}

\newcommand{\rsun}{\mathrm{R_{\odot}}} 
\newcommand{\Rhe}{R_{\mathrm{He}}} 
\newcommand{\Rwd}{R_{\mathrm{WD}}} 

\newcommand{\iniMwd}{M^{i}_{\mathrm{WD}}}

\newcommand{\iniTwd}{T^{i}_{\mathrm{c},\mathrm{WD}}}
\newcommand{\iniP}{ P^{i}_{\mathrm{orb}}}

\newcommand{\iniSc}{s^{i}_{\rm c}}

\newcommand{\mesa}{{\tt\string MESA}}

\newcommand{\aesopus}{\AE SOPUS}

\newcommand{\kB}{k_{\mathrm{B}}}
\newcommand{\NA}{N_{\mathrm{A}}}

\newcommand{\cp}{c_{\mathrm{p}}}

\newcommand{\taumdot}{\tau_{\dot{M}}}
\newcommand{\tauth}{\tau_{\mathrm{th}}}

\newcommand{\taugr}{\tau_{\mathrm{gr}}}

\newcommand{\tauconv}{\tau_{\mathrm{conv}}}
\newcommand{\tauconvsol}{\tau_{\mathrm{conv},\odot}}

\newcommand{\Jorb}{J_{\mathrm{orb}}}
\newcommand{\Jdot}{\dot{J}}
\newcommand{\Jdotgr}{\dot{J}_{\mathrm{GWR}}}
\newcommand{\Jdotmb}{\dot{J}_{\mathrm{MB}}}

\newcommand{\JdotRVJ}{\dot{J}_{\mathrm{RVJ}}}
\newcommand{\JdotSills}{\dot{J}_{\mathrm{S00}}}
\newcommand{\JdotMatt}{\dot{J}_{\mathrm{M15}}}
\newcommand{\JdotCARB}{\dot{J}_{\mathrm{CARB}}}

\newcommand{\Jdotunit}{\rm{g}\,\rm{cm}^{2}\,\rm{s}^{-2}}

\newcommand{\Porb}{P_{\mathrm{orb}}}
\newcommand{\Prot}{P_{\mathrm{rot}}}
\newcommand{\Pcrit}{P_{\mathrm{crit}}}
\newcommand{\Psol}{P_{\odot}}

\newcommand{\Pdot}{\dot{P}}
\newcommand{\pdotunit}{\rm{s}\,\rm{s}^{-1}}

\newcommand{\Firr}{F_{\rm{irr}}}

\newcommand{\Sc}{s_{c}}

\newcommand{\alfv}{Alfv\'en~}
\newcommand{\Ro}{Ro}
\newcommand{\Rocrit}{Ro_{\rm crit}}

\defcitealias{Wong2021}{WB21}
\defcitealias{Bildsten2006}{B06}
\defcitealias{vanRoestel2022}{vR22}
\defcitealias{Ferguson2005}{FA05}

\defcitealias{Matt2015}{M15}
\defcitealias{Van2019}{CARB}
\defcitealias{Rappaport1983}{RVJ}
\defcitealias{Sills2000}{S00}

\newcommand{\rvj}{\citetalias{Rappaport1983}~MB~}
\newcommand{\sills}{\citetalias{Sills2000}~MB~}
\newcommand{\carb}{\citetalias{Van2019}~MB~}
\newcommand{\matt}{\citetalias{Matt2015}~MB~}

\begin{document}

\title{Accretor Luminosity as Probe of Orbital Evolution and Magnetic Braking in AM~CVn Binaries}

\author[0000-0001-9195-7390,sname='Wong',gname='Tin Long Sunny']{Tin Long Sunny Wong}
\affiliation{The Observatories of the Carnegie Institution for Science, Pasadena, CA 91101, USA}
\email[show]{swong@carnegiescience.edu}

\author[0000-0003-4221-9097]{Peter Scherbak}
\affiliation{TAPIR, California Institute of Technology, Pasadena, CA 91125, USA}
\affiliation{Astronomy Department, California Institute of Technology, Pasadena, CA 91125, USA}
\email{pscherba@astro.caltech.edu}

\author[0000-0002-4544-0750]{Jim Fuller}
\affiliation{TAPIR, California Institute of Technology, Pasadena, CA 91125, USA}
\email{jfuller@caltech.edu}

\correspondingauthor{Tin Long Sunny Wong}
\email{swong@carnegiescience.edu}

\begin{abstract}

AM~CVn binaries are ultracompact systems with orbital periods $P_{\rm orb} \lesssim 70$~min, where a white dwarf accretes from a He-rich semidegenerate donor. 
At short orbital periods ($P_{\rm orb} \lesssim 20$~min), rapid accretion reheats the accretor. 
As the system widens and the mass transfer rate decreases, the accretor follows standard white dwarf cooling at longer orbital periods ($P_{\rm orb} \gtrsim 30$~min). 
The accretor cooling luminosity then acts as a powerful probe into the orbital evolution of the binary. 
Recent observations show that the accretor appears brighter (younger) at long orbital periods, compared 
to models where orbital angular momentum loss is solely set by gravitational wave emission. 
We show that additional angular momentum loss, which we ascribe to magnetic braking of the donor, 
can better reproduce observations of accretor luminosity. We test and constrain various proposed 
magnetic braking formalisms, for the white dwarf, helium star, and evolved cataclysmic variable (CV) channels for AM~CVn formation. 
None of the magnetic braking formalisms in their original form can reproduce the observations, but 
the widely popular magnetic braking proposed by Rappaport et al. 1983 boosted by a factor $\approx 5$, 
or the saturated magnetic braking proposed by Matt et al. 2015 boosted by a factor of $\approx 150$, 
yields satisfactory results. 
We also find that the CARB formalism, recently applied to the evolved CV channel, is strongly in tension with observations of accretor luminosity. 
Our models motivate UV spectroscopy observations, and dedicated WD atmosphere models for AM~CVn accretors.

\end{abstract}


\section{Introduction}

AM Canum Venaticorum (AM~CVn) binaries are semidetached systems with a white dwarf (WD) accretor and a semidegenerate helium (He)-rich donor, residing at orbital periods ($\Porb$) between 5 and 68 min \citep[][]{Ramsay2018,Green2020}. 
For recent comprehensive literature reviews, we refer to  \cite{Green2025}. 
Systems at the shortest periods are strong gravitational wave sources in the millihertz frequency \citep[e.g.,][]{Nelemans2004,Breivik2018,Kupfer2024} detectable by future space-based gravitational wave detectors such as Laser Interferometer Space Antenna \citep[LISA;][]{AmaroSeoane2017}, TianQin \citep[][]{Luo2016}, etc. 
At $\Porb \lesssim 10$~min, rapid accretion of He at $\mdot \sim 10^{-7} - 10^{-8} \, \msunyr$ 
leads to strong compressional heating of the accretor's He envelope. A thermonuclear runaway can ensue \citep[e.g.,][]{Nomoto1982,Shen2009}, forming transients such as He novae \citep[e.g.,][]{Woudt2009}, type Ia supernovae via the double detonation mechanism \citep[e.g.,][]{Woosley2011}, .Ia supernovae \citep[e.g.,][]{Bildsten2007}, and calcium-strong transients \citep[e.g.,][]{Perets2010,Kasliwal2012,Hall2026}.

There are three formation channels for AM~CVn binaries:
(a) the WD channel, where the donor is a He WD formed from a common envelope event \citep[e.g.,][]{Paczynski1967,Webbink1984,Nelemans2001,Chen2022}, 
(b) the He star channel, where the donor is formerly a He-burning star that has now ceased burning and begun to cool due to mass transfer \citep[e.g.,][]{Savonije1986,Iben1987,Yungelson2008,Bauer2021,Sarkar2023_HeStar,Rajamuthukumar2025}, 
(c) the evolved cataclysmic variable (CV) channel \citep[e.g.,][]{Tutukov1985,Podsiadlowski2003,Goliasch2015,Liu2021}, where the donor star undergoes stable mass transfer, from a H-rich phase as an evolved star with a He core, and later in a He-rich phase once the He core is exposed. 

Distinguishing between the three formation channels has proven difficult, since they more-or-less yield similar predictions for the mass-radius relation of AM~CVn donors. Abundance measurements of the accreted material shows promise in breaking the degeneracy \citep[e.g.,][]{Nelemans2010,Rodriguez2023,Yu2026,Yu2026b}. 

Our paper covers all three formation channels, and addresses observations that 
the optical luminosities of AM~CVn binaries exhibit a decreasing trend with $\Porb$ \citep[][]{Ramsay2018,Green2025}. 
This trend qualitatively agrees with previous theoretical works assuming that the orbital angular momentum loss (hereafter AML) is solely driven by gravitational wave radiation (hereafter GWR; see \citealt{Bildsten2006}, hereafter \citetalias{Bildsten2006}, \citealt{Wong2021}, hereafter \citetalias{Wong2021}). 
At short $\Porb$, the high $\mdot$ leads to the system luminosity being dominated by accretion, while compressional heating reheats the accretor and resets its cooling age (\citetalias{Bildsten2006}). 
As the donor loses mass and the orbit widens to longer $\Porb$, $\mdot$ declines. 
Eclipse studies show that at $\Porb \gtrsim 50$~min, the optical luminosity of the system is mostly dominated by the accretor (\citealt{Green2018}, \citetalias{vanRoestel2022}, \citealt{Rodriguez2023}). 
At this point, the accretor evolves as a cooling WD, with its cooling age encoded in $\Porb$ (\citetalias{Bildsten2006}, \citetalias{Wong2021}). 
These long-period systems are therefore excellent testbeds for the orbital evolution. 

However, observations of long-period AM~CVn binaries show that theory underpredicts their optical luminosities at fixed $\Porb$ (\citealt{Ramsay2018}, \citetalias{vanRoestel2022}, \citealt{Macrie2024}, \citealt{Householder2026}). 
Since eclipse studies show little to no influence by the accretion disk and hot spot in the optical (\citetalias{vanRoestel2022}), this suggests that the binary experiences additional AML, 
which speeds up the orbital evolution, such that the accretor is younger, and hence brighter, at a given $\Porb$. 

One potential source of additional AML is magnetic braking (MB) of the donor, as previously applied by \cite{Belloni2023} to the evolved CV channel. 
If the donor loses mass via a magnetized wind that corotates with the stellar surface out to the \alfv radius, the much larger lever arm can efficiently extract angular momentum from the stellar spin \citep[e.g.,][]{Weber1967,Mestel1987}. This in turn results in AML from the orbit, since the donor is likely tidally synchronized \citep[e.g.,][]{Fuller2012,Fuller2013}. 
Adopting the Convection and Rotation-Boosted (CARB) MB formalism \citep[][]{Van2019}, \cite{Belloni2023} showed that enhanced AML eliminates the fine-tuning problem for AM~CVn formation \citep[see also][]{Tang2026,Zhao2026}, and the donor remains inflated with surface hydrogen at undetectable levels (mass fraction $X \lesssim 10^{-6}$). Nevertheless, how this manifests in the accretor luminosity evolution, and its application beyond the evolved CV channel, remains untested. 
Furthermore, given the great uncertainty in the strength of MB in binaries \citep[e.g.,][]{ElBadry2022,Belloni2024}, not to mention single stars \citep[e.g.,][]{vanSaders2016}, a test of the various proposed MB formalisms and strengths is desirable. 

In this paper, we test various forms of MB and their influence on the accretor luminosity evolution. 
Our paper is organized as follows. 
In Section~\ref{sec:numerics}, we describe our binary evolution models, adopting the WD channel as our fiducial. 
We show in Section~\ref{sec:GWR only} that GWR-only models underpredict the accretor luminosity, and discuss theoretical uncertainties related to irradiation of the donor and low-temperature radiative opacities, which marginally mitigate the problem. 
In Section~\ref{sec:MB}, we show that MB leads to a faster orbital evolution, and hence substantial improvement to the models. While none of the four tested MB prescriptions yield satisfactory results, we varied the MB strength and found empirically the AML required to reproduce observations. We show in Section~\ref{sec:alternative formation channels} that our results hold even with the evolved CV and He star channels. In Section~\ref{sec:implications}, we discuss observational implications for period derivatives, X-ray luminosities, and period distributions. 
We conclude in Section~\ref{sec:conclusion}.

\section{Numerical Setup}
\label{sec:numerics}

We evolve AM~CVn binaries with the stellar evolution code Modules for Experiments in Stellar Astrophysics \citep[$\mesa$;][]{MESAI,MESAII,MESAIII,MESAIV,MESAV,MESAVI}. 
We first evolve the donor along with the orbit, with the $\mesa$ version \verb|r12778| binary module, where we do not resolve the structure of the accretor and only evolve its mass. 
Then, given the $\mdot$ tracks, we model the accretor's detailed stellar structure upon mass accretion with the $\mesa$ version \verb|24.08.1| star module, without re-evolving the binary. 
This separation is done largely due to the numerical difficulty of evolving the donor in $\mesa$ versions that follow \verb|r12778|, and the inclusion of better physics related to crystallization and phase separation in \verb|24.08.1| \citep[][]{Jermyn2021,Bauer2023}, but is entirely inconsequential because the detailed structure of the accretor does not affect the orbital evolution. 

The $\mesa$ EOS is a blend of the OPAL \citep{Rogers2002}, SCVH
\citep{Saumon1995}, FreeEOS \citep{Irwin2004}, HELM \citep{Timmes2000},
PC \citep{Potekhin2010}, and Skye \citep{Jermyn2021} EOSes.
Radiative opacities are primarily from OPAL \citep{Iglesias1993,
Iglesias1996}, with the high-temperature, Compton-scattering dominated regime by
\citet{Poutanen2017}. Electron conduction opacities are from
\citet{Cassisi2007} and \citet{Blouin2020}.
Nuclear reaction rates are from JINA REACLIB \citep{Cyburt2010}, NACRE \citep{Angulo1999} and
additional tabulated weak reaction rates \citet{Fuller1985, Oda1994,
Langanke2000}.  Screening is included via the prescription of \citet{Chugunov2007}.
Thermal neutrino loss rates are from \citet{Itoh1996}. 
Our $\mesa$ input and output files are publicly available at \href{https://doi.org/10.5281/zenodo.24357432}{doi.org/10.5281/zenodo.24357432}. 

\subsection{Initial stellar models}

Early works on AM~CVn binaries assumed a fully degenerate (hereafter low-entropy) He WD, but \citetalias{Bildsten2006}, \cite{Deloye2007} \& \citetalias{Wong2021} showed that the donor may be partially degenerate (hereafter high-entropy), particularly if it comes into contact very shortly after formation, and hence has little time to cool \citep[$\lesssim 10^{8}$~yrs; see][]{Wong2023}. Recent measurements of eclipsing AM~CVn binaries (\citealt{Green2018}, \citetalias{vanRoestel2022}) revealing that the donors have inflated radii, and the discovery of a 13.7 min eclipsing system with a warm $\approx 16000$~K He-rich donor \citep[][]{Burdge2023}, are both consistent with the high-entropy He WD picture. 

Motivated by these observations, we take a $0.15\,\msun$ He WD, with various values of entropies, as our initial donor. 
We created the initial donor following \citetalias{Wong2021} and the ${\mesa}$ {\verb|make_he_wd|} test suite. We evolve a $1.5\,\msun$ star up to the formation of a $0.15\,\msun$ He core, where we use a H mass fraction $X = 10^{-5}$ to define the core-envelope boundary. The H envelope is then removed via a wind. This procedure implicitly assumes that the binary survives any H-rich mass transfer phase prior to the AM~CVn phase \citep[e.g.,][however, see \citealt{Shen2015}]{Kaplan2012}. 
The newly formed He WD is allowed to cool to central specific entropy values of $\iniSc = 2.7 - 3.8 \, \kB \NA$, where $\kB$ is Boltzmann's constant and $\NA$ is the Avogadro number. In the rest of this paper, we label our models by their initial central specific entropies $\iniSc$.

The CO WD accretors are created similarly following the $\mesa$ \verb|make_co_wd| test suite. They are allowed to cool to a central temperature of $10^{7}$~K. As \citetalias{Wong2021} showed, the initial central temperature of the accretor does not matter much at long $\Porb$, as long as the accretion reheating happens. 
Our base model assumes an initial accretor mass of $\iniMwd = 0.7\,\msun$, such that its final mass is $\approx 0.8 - 0.85\,\msun$ after accreting most of the donor's mass; eclipse studies show that the accretor mass is $\approx 0.8 - 0.9\,\msun$ (e.g., \citealt{Green2018}, \citetalias{vanRoestel2022}, \citealt{Rodriguez2023}, \citealt{Green2025}). 

\subsection{Binary evolution}

{\mesa} models the stellar structure of the donor, computes the secular mass transfer rate $\mdot$ using the \cite{Ritter1988} prescription accounting for sources and sinks of orbital angular momentum, and finally evolves the binary component masses and separation. 
In this work, we assume fully conservative mass transfer (however, see Footnote~\ref{footnote5}). 
We do not model the spins of the stars, since their spins are small compared to the orbital angular momentum. 
Tidal synchronization of the donor is implicitly assumed in the MB models in Sections~\ref{sec:MB} \& \ref{sec:alternative formation channels}, which is reasonable since tidal synchronization is expected to be efficient for stars filling their Roche lobes. 
Since our donors are generally inflated, the wider binary separation allows the binary to avoid direct impact accretion which tends to destabilize mass transfer \citep[e.g.,][]{Marsh2004,Gokhale2007}, justifying our neglect of the spins (see also \citetalias{Wong2021}, \citealt{McNeill2025}). 
We initiate the binary such that the donor is Roche-lobe-filling. This is to prevent any cooling prior to mass transfer. 
Our base model assumes that the sole mechanism for AML comes from GWR (Section~\ref{sec:GWR only}), and in Section~\ref{sec:MB}, we include the effects of MB. 
Our binary models terminate once anywhere in the donor reaches $\log_{10} Q = 5$, where $\log_{10} Q = \log_{10} \rho - 2 \log_{10} T + 12$, above which $\mesa$ has no EOS coverage. 

In Section~\ref{subsec:irradiation}, we explore the effects of irradiation on the donor following \cite{Deloye2007}. We show that it changes the thermal evolution of the donor and consequently the orbital evolution. We calculate the accretion luminosity $\Lacc = G \mwd \mdot / \Rwd $ from the secular $\mdot$, where $G$ is the gravitational constant, $\mwd$ and $\Rwd$ are the accretor mass and radius assuming a cold WD mass-radius relation \citep[][]{Verbunt1988}. 
The irradiation flux received by the donor is then $\Firr = \eta \Lacc ( \Rhe / 2 a )^{2}$, where $\Rhe$ is the donor radius, $a$ is the binary separation, and we assume some variable efficiency factor $\eta \approx 0.1 - 0.5$. These values may be optimistic, but our goal is to show the effects of irradiation. 
For numerical convergence, we cap $\Firr$ at a maximum value of $10^{12}$~erg~cm$^{-2}$~s$^{-1}$, which is only inaccurate near the period minimum. 
For simplicity, this flux is uniformly deposited at a column depth $\Sigma = 1$~g~cm$^{-2}$ \citep[for implementation, see][]{MESAII}. 

\subsection{Adiabatic evolution of donor}
\label{subsec:adiabatic}

We calculated evolutionary tracks in the limit that the donor evolves adiabatically (see also \citetalias{Bildsten2006}), where mass transfer is so rapid that the donor is unable to cool, as follows. The adiabatic mass-radius tracks are obtained from $\mesa$ binary runs where we set an arbitrarily large AML (and hence high $\mdot$), such that the central specific entropy of the donor remains constant by definition. Given the adiabatic mass-radius relation, we calculated $\mdot$ semi-analytically \citep[e.g.,][]{Brooks2015,Bauer2021}:
\begin{equation}
    \frac{\mdot}{\mhe} \left( \frac{ \xi }{2} + \frac{5}{6} - \frac{\mhe}{\mwd} \right) = \frac{\Jdotgr}{\Jorb} ,
\end{equation}
where $ \xi = d \ln \Rhe / d \ln \mhe$ is taken from the adiabatic mass-radius relation, $\Jdotgr$ is the torque due to GWR \citep[][]{Landau1975}, and $\Jorb$ is the orbital angular momentum. For a fully degenerate WD, $\xi = -1/3$, but realistic values range from $-0.2$ to $-0.1$ (see, e.g., \citetalias{Wong2021}).

\subsection{Donor opacity}
\label{subsec:donor opacity}

The microphysics of our donor models generally follows \citetalias{Wong2021}, except for the low-temperature radiative opacities. 
In \citetalias{Wong2021}, the radiative opacities at $\log_{10}(T/\rm{K}) \lesssim 3.8$ are taken from \cite{Ferguson2005}, for a hydrogen-free, solar abundance mixture, which is limited to $\log_{10} R \leqslant 1$, where $\log_{10} R = \log_{10} \rho - 3 \log_{10} T + 18 $. 
At $\log_{10} R > 1$, $\mesa$ extrapolates by taking the value at $\log_{10} R = 1$ and the same $\log_{10} T$. As noted in \citetalias{Wong2021}, at long $\Porb$, the top radiative-convective boundary of the donor enters this extrapolated regime. 
Since the donor's surface boundary condition and intrinsic luminosity depend on the opacity, the donor's cooling and hence mass-radius evolution are uncertain. In turn, the mass transfer rate $\mdot$ and hence orbital evolution are impacted. 

In this work, we created low-temperature, grain-free radiative opacity tables using the \aesopus~2.1~web interface \citep[][]{Marigo2009,Marigo2022,Marigo2024}. 
We specify a hydrogen-free composition with base metallicity equal to solar, $Z = 0.0142$, and abundances given by \cite{Asplund2009}, and additionally enhanced nitrogen and depleted carbon and oxygen, suitable for a post-CNO-burning He core. 
Our tables cover $-8 \leqslant \log_{10} R \leqslant 6$ and $3 \leqslant \log_{10} T \leqslant 4.5$, where the broader range in $\log_{10}R$ is thanks to the latest extension of \aesopus~to high pressure environments intended for use in brown dwarfs and planets \citep[][]{Marigo2024}. Certain high-density effects are still missing, such as He collisional induced absorption \citep[e.g.,][]{Kowalski2014,Blouin2024} and high-density corrections to the He-minus free-free absorption and He Rayleigh scattering \citep[e.g.,][]{Iglesias2002,Rohrmann2018}. 
Inclusion of He collisional induced absorption increases the opacity, as does inclusion of pressure ionization which provides more free electrons and enhances He-minus free-free absorption. However, correlations between helium atoms reduce the absorption coefficients of He-minus free-free absorption and He Rayleigh scattering by as much as an order-of-magnitude \citep[][]{Iglesias2002}. The overall effect of these high-density effects is therefore unclear, and should be considered in future studies. 

While the surface radiative-convective boundary of the donor still enters regimes not covered by our tables (in which case an extrapolation is similarly done), the extent is much more limited, and we proceed with these cautions in mind. 
We refer to Appendix~\ref{sec:appendix opacity table} for a comparison of the opacities used in \citetalias{Wong2021} and in this work. 
Our \aesopus~tables are included in our Zenodo repository, with the hope of them being useful to the community.

\section{Evolution with only GWR}
\label{sec:GWR only}

\begin{figure*}[t!]
\centering
\fig{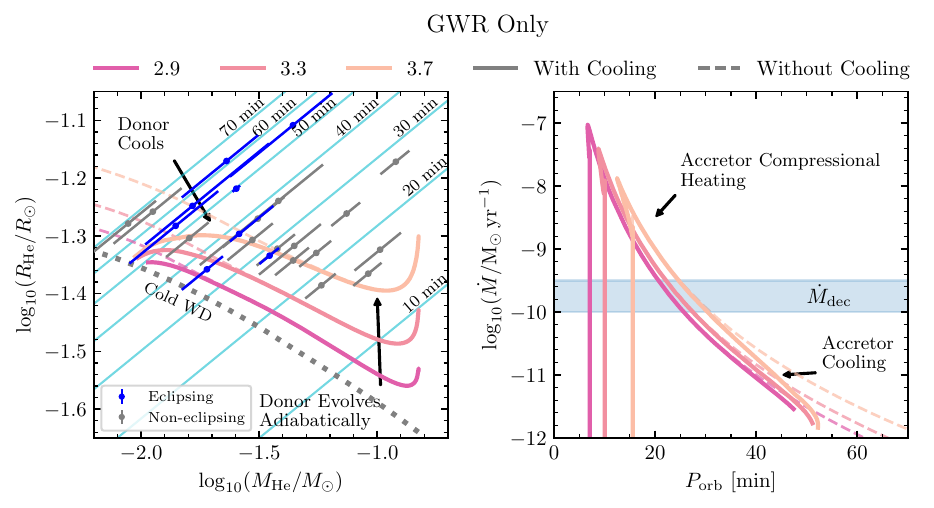}{ \textwidth }{}
\vspace{-20 pt}
\caption{ 
Mass-radius relation of the donor (left) and evolution of the mass transfer rate (right), wherein the only orbital angular momentum loss is from gravitational wave radiation. 
Solid lines show donors evolved with cooling allowed, and dashed without (i.e., adiabatic). 
The models are labeled by their initial central specific entropy values $\iniSc$ in units of $\kB \NA$. 
\textit{Left:} The general evolution is from bottom right to top left. The fully degenerate solution is shown in the grey dotted line. The diagonal light blue lines show various fixed $\Porb$. 
Observations are shown in the error bars, with eclipsing systems in dark blue and non-eclipsing in grey. 
\textit{Right:} The general evolution is from left to right. The blue shade denotes the decoupling mass transfer rate $\mdotdec$, above which the accretor luminosity is set by compressional heating, and below which the accretor simply cools. 
\label{fig:MR_and_Mdot_GW}}
\end{figure*}

In this section, we present the evolution of the binary where the only AML is from GWR. A detailed exploration is presented in \citetalias{Wong2021}, and here we elucidate some key features in subsection \ref{subsec:general features}.

\subsection{General Features}
\label{subsec:general features}

\subsubsection{Donor Evolution}
\label{subsubsec:GWR donor evolution}

We show the mass-radius evolution of the donor, and the evolution of $\mdot$, in Figure~\ref{fig:MR_and_Mdot_GW}. 
We refer to \cite{Deloye2007} for a detailed description of the donor evolution. 
As the donor loses mass, it initially contracts as its outermost high-entropy envelope is removed. 
The orbit shrinks to shorter $\Porb$, and $\mdot$ increases. 
Eventually, the expansion of the inner degenerate layers dominates the mass-radius relation of the donor, and the donor begins to expand as it loses mass. 
The binary reaches period minimum and later widens to longer $\Porb$, as $\mdot$ peaks and subsequently decreases. 
This is due to the longer GW inspiral time, $\taugr$, at longer $\Porb$ \citep[for an in-depth discussion, see][]{Bauer2021}. 

The thermal evolution of the donor is dictated by two timescales: the mass transfer timescale, 
\begin{align}
\label{eqn:tau_mdot}
    \taumdot = \frac{ \mhe }{ \dot{M} }, 
\end{align}
and the thermal timescale of the donor, 
\begin{align}
\label{eqn:tau_th}
    \tauth = \frac{ \int^{\mhe}_{0} \cp T dm }{ \Lhe }, 
\end{align}
where $\cp$ is the specific heat capacity and $\Lhe$ is the intrinsic luminosity of the donor. 

At short $\Porb$, the rapid mass loss leads to $\taumdot < \tauth$. There is no time for the donor to radiate its heat content, and the evolution of the donor is adiabatic; its central entropy remains roughly constant $\approx \iniSc$ with time. Therefore, the mass-radius relation of the donor closely follows that of the adiabatic line, along with the associated $\mdot$ (see Figure~\ref{fig:MR_and_Mdot_GW}). 

At longer $\Porb$, there comes a time when $\mdot$ is low enough that $\taumdot \gtrsim \tauth$. The donor can now lose entropy and shrink towards the fully degenerate solution, as seen in Figure~\ref{fig:MR_and_Mdot_GW} at $\Porb \gtrsim 40$~min. However, the $\Porb$ at which this happens, depends on the opacity at the surface radiative-convective boundary of the donor, as we will illustrate in Subsection~\ref{subsec:opacity}. 

As Figure~\ref{fig:MR_and_Mdot_GW} shows, a higher-entropy donor has a larger radius at a given mass. This in turn means that it reaches period minimum at a longer $\Porb$, with a lower peak $\mdot$. 
Furthermore, it has a larger mass at a given $\Porb$, which in turn leads to a larger $\Jdotgr$ and hence $\mdot$, as seen in Figure~\ref{fig:MR_and_Mdot_GW}. In other words, a higher-entropy donor has a more rapid orbital evolution. 
Nevertheless, once it cools, the orbital evolution slows. 
Both the initial thermal content of the donor, and its subsequent evolution, have important implications for the luminosity evolution of the accretor. 

In the left panel of Figure~\ref{fig:MR_and_Mdot_GW}, we also show the inferred masses and radii of observed systems, taken from \cite{Green2018}, \citetalias{vanRoestel2022}, and \cite{Rodriguez2023} for eclipsing systems, and from the \cite{Green2025} catalog (and references therein) for the spectroscopic and superhump methods. The uncertainties are correlated, because the $\Porb$ of a Roche-lobe-filling donor is determined by its average density \citep[e.g.,][]{Faulkner1972}. 
We note that for the superhump systems, the mass ratio $q$ is derived using the superhump period \citep[e.g.,][]{Patterson2005,Kato2013,McAllister2019}, and we assume an accretor mass $\mwd = 0.8 \pm 0.1\,\msun$ motivated by findings from eclipse studies. 

As already put forward in \cite{Roelofs2006_AMCVN}, \cite{Green2018} and \citetalias{vanRoestel2022}, observations are consistent with a high-entropy donor. Not only are they born with high entropy, but they also have to remain hot at long $\Porb$. In comparison, our fiducial models cool too soon. 
As we will show, irradiation \citep[e.g.,][]{Deloye2007} and some extra AML \citep[e.g.,][]{Belloni2023} can help alleviate this problem. 

\subsubsection{Accretor Evolution}
\label{subsubsec:GWR accretor evolution}

\begin{figure*}[t!]
\centering
\fig{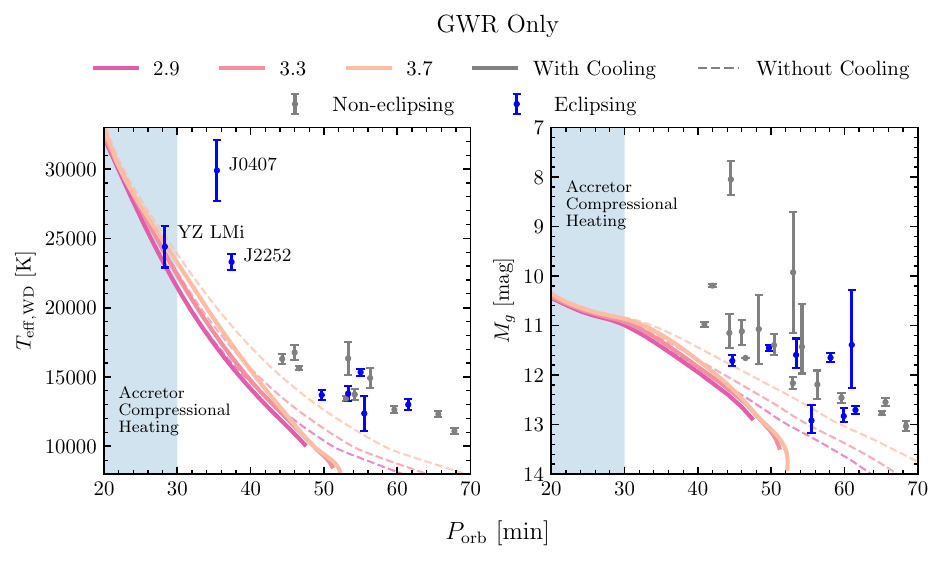}{ \textwidth }{}
\vspace{-20 pt}
\caption{ 
Accretor effective temperature (left) and Pan-STARRS1 $g$-band absolute magnitude (right) of our models, as presented in Fig.~\ref{fig:MR_and_Mdot_GW},  evolving to the right towards longer $\Porb$. Solid lines show models where the donor is allowed to cool, and dashed lines show models with the donor evolving adiabatically. 
The colors of the models follow their initial central specific entropy values $\iniSc$ (in units of $\kB \NA$) in Fig.~\ref{fig:MR_and_Mdot_GW}. 
The errorbars show inferred values from observations (see text), with dark blue denoting eclipsing systems and grey denoting non-eclipsing systems. The shaded region ($\lesssim 30$~min) shows where compressional heating is expected to dominate the accretor luminosity. 
\label{fig:Teff_and_Mg_GW}}
\end{figure*}

We now describe the evolution of the accretor. We refer to \citetalias{Bildsten2006} for a pioneering study. 
At short $\Porb$, the high $\mdot$ leads to strong compression in the He envelope of the accretor, which drives a temperature inversion\footnote{Our models avoid the occurrence of a He flash due to the lower peak $\mdot$, but see \cite{Wong2023} for an exploration of this, in a different part of the parameter space.}. Heat flows down the temperature gradient both towards the surface and towards the cool core. During this stage, the surface luminosity is set by compressional heating and is dependent on $\mdot$ (e.g., \citealt{Townsley2004}, \citetalias{Bildsten2006}). 
After roughly a thermal conduction time $\approx 10^{7}$~yrs, the heat front reaches the center. 
The accretor is now reheated, and its cooling age is reset. 

As $\mdot$ declines, eventually the surface luminosity of the accretor decouples from $\mdot$, since compressional heating is inefficient compared to the intrinsic cooling of the accretor. \citetalias{Bildsten2006} showed that this decoupling occurs around $\mdotdec \approx (1-3)\times 10^{-10}\,\msun$. 
At $\mdot \lesssim \mdotdec$, which occurs at $\Porb \gtrsim 30$~min as labeled in Figure~\ref{fig:MR_and_Mdot_GW}, the surface luminosity of the accretor is simply that of a cooling WD, $\Lwd \propto t^{-5/3}$ \citep[][]{Townsley2004}, where $t$ is the cooling time. One difference is that compared to a normal CO WD, the cooling luminosity is slightly higher due to the higher specific heat capacity of the thick He shell (\citetalias{Wong2021}).

The left panel of Fig.~\ref{fig:Teff_and_Mg_GW} shows the evolution of the accretor effective temperature, $\Teffwd$, given different mass transfer histories set by the donor. 
The key features are (i) at short $\Porb \lesssim 30$~min, $\Teffwd$ is set by compressional heating, (ii) at long $\Porb \gtrsim 30$~min, $\Teffwd$ is set by WD cooling, and (iii) binaries with faster orbital evolution have a hotter accretor at a given $\Porb$. 

At long $\Porb$, accretors with higher-entropy donors are hotter: this is a result of the faster orbital evolution due to the more massive donor. 
However, once the donor cools, the binary spends significant time without much advancing in $\Porb$, while the accretor continues to cool. This manifests in a downturn in $\Teffwd$ and $M_{g}$ at long $\Porb$. 
Given the uncertainties in the cooling of the donor (see Sec~\ref{subsec:irradiation}~\&~\ref{subsec:opacity}), we also computed evolutionary tracks where the donors evolve adiabatically, i.e. they are not allowed to cool (see Sec~\ref{subsec:adiabatic}). 
At a given $\Porb$, because the donor is more massive, the orbital evolution is faster, and the accretor is hotter. 


\subsubsection{Comparison to Observations}
\label{subsubsec:GWR comparison to observations}

We compare our predicted $\Teffwd$ with those inferred by \cite{Macrie2024} from blackbody fits to the spectral energy distribution (SED). 
We choose to compare with systems with $\Porb \geqslant 40$~min, since the SEDs of short $\Porb$ systems may still have significant contributions from the quiescent accretion disk \citep[e.g.,][]{Kara2025}. 
We reexamined the fits using VOSA \citep[][]{VOSA}, and found general agreement with \cite{Macrie2024}. 
We removed the system ASASSN-21au since we are not able to retrieve its parallax from the Gaia~DR3 catalogue. 
We found considerable differences for two systems and opted to adopt the VOSA values: we adopted $\Teffwd = 14900\pm700$~K instead of $19570\pm1440$~K for SDSSJ1552+3201, and $13800\pm600$~K instead of $16400\pm550$ for ZTFJ0220+2141. We favor these adopted values because they agree better with the findings of \cite{Roelofs2007_J1552} and \cite{vanRoestel2022}, respectively. 
We checked that the inferred $\Teffwd$ tends to be slightly lower by $\approx 1000$~K when fitted to pure-He DB WD model atmospheres using WDPhotTools \citep[][]{WDPhotTools}, which is similar to the findings of \cite{Rau2010}, but this does not change our conclusions much. 
We also show the 28, 35 and 37-min systems, YZ~LMi, ZTF~J0407-0007 and ZTF~J2252-0519, which have $\Teffwd = 24400\pm1500$, $29900\pm2200$ and $23300\pm600$~K respectively \citep[][]{Yu2026,Yu2026b}. These are measured from Hubble Space Telescope (HST) UV spectroscopy accounting for contamination by the accretion disk/hotspot, using detailed WD atmosphere models that allow for varying elemental abundances. 
The blackbody fit by \cite{Macrie2024} gives $15560\pm460$~K for ZTF~J2252-0519, which highlights the problems of photometrically derived temperatures, particularly at short $\Porb$. 

In the right panel of Fig~\ref{fig:Teff_and_Mg_GW}, we compare synthetic Pan-STARRS \citep[][]{PANSTARRS} $g$-band magnitudes of our models to observations, similarly to \cite{Ramsay2018}. 
We chose this band because it minimizes contamination from other sources, such as the accretion bright spot at shorter wavelengths \citep[e.g.,][]{Yu2026} and the accretion disk at longer wavelengths (e.g., \citetalias{vanRoestel2022}, \citealt{ElBadry2026}). 
We took $\Porb \geqslant 40$~min systems from the \cite{Green2025} catalogue, and queried Vizier for their Pan-STARRS magnitudes (for ASASSN-14ei we adopted Dark Energy Survey photometry, and for SDSS~J0804+1616, Sloan Digital Sky Survey photometry). We checked their light curves in the Zwicky Transient Facility \citep[][]{Graham2019} to ensure that the magnitudes correspond to quiescence. We corrected for reddening, using the G-Tomo 3D dust maps to obtain $A_{\rm V}$ \citep[][]{Lallement2022,Vergely2022}, and adopting $A_{g} = 1.1 A_{\rm V}$ \citep[][]{Cardelli1989}. Finally, we computed the absolute magnitudes using the Gaia DR3 geometric distances from \cite{BailerJones2021} as compiled by \cite{Green2025}. 
In addition, we included three eclipsing systems with reliable Gaia parallaxes from the recent ZTF discoveries by \cite{Householder2026} and \cite{vanRoestel2026}.
We interpolated the pure-He DB~WD atmosphere models by \cite{Bergeron2011}\footnote{\url{http://www.astro.umontreal.ca/~bergeron/CoolingModels}}, to obtain synthetic photometry from our $\mesa$ models.

\textit{Both comparisons show that the observed systems appear hotter and brighter than predicted}, as pointed out in \cite{Ramsay2018}, \citetalias{vanRoestel2022}, \cite{Macrie2024}, \cite{Green2025}, \cite{Yu2026b} and \cite{Householder2026}. 
This is true even in the most optimistic case, where the donor evolves adiabatically. The problem worsens with having a low-entropy donor and allowing the donor to cool. 
Furthermore, it is unclear what allows the donor evolution to be adiabatic even at long $\Porb$. We will show in Section~\ref{subsec:irradiation} that irradiation effects bring the donor evolution closer to adiabatic, but does not fix the problem completely. 

\begin{figure*}[t!]
\centering
\fig{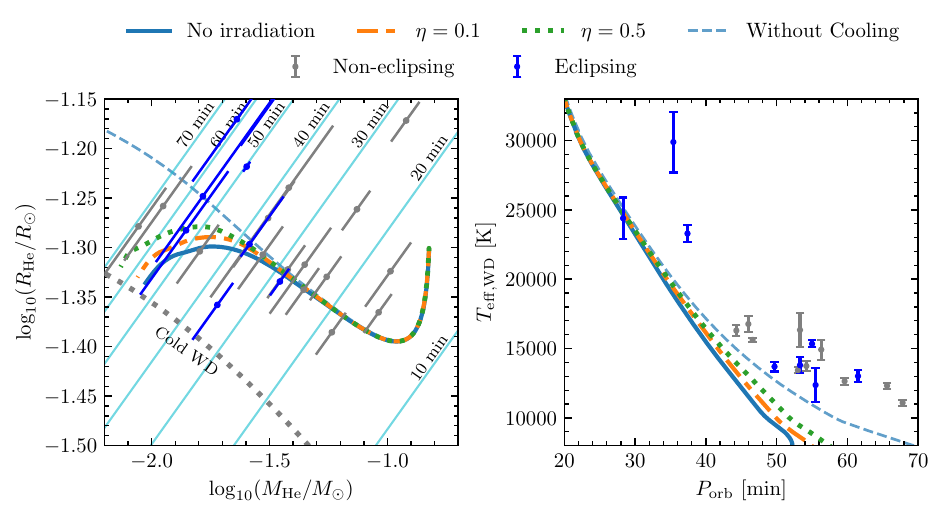}{ \textwidth }{}
\vspace{-25 pt}
\caption{ 
Effects of irradiation on the donor mass-radius relation (left) and accretor effective temperature (right) of our models with a $\iniSc = 3.7\,\kB \NA$ donor. Solid blue lines show models without irradiation (as presented in Sec.~\ref{subsec:general features}), dashed orange lines and dotted green lines show models with irradiation efficiency $\eta = 0.1$ and $0.5$, respectively, and dashed blue lines show models where the donor evolves adiabatically. 
Data and other annotations same as Fig.~\ref{fig:MR_and_Mdot_GW} \& \ref{fig:Teff_and_Mg_GW}. 
\label{fig:irradiation}}
\end{figure*}

However, these comparisons come with their caveats \citep[as raised by][]{Rodriguez2023}. The inferred $\Teffwd$ assume a blackbody SED, and the DB~WD atmosphere models assume a pure-He surface. Neither may correspond to the SED of a He-rich atmosphere polluted with metals at a level much higher than known DBZ WDs, nor do they account for possible contamination by the quiescent accretion disk and bright spot (though, the contamination in the optical appears minimal in long-$\Porb$ eclipsing systems; \citetalias{vanRoestel2022}). 
We also caution that observations may be biased towards outbursting and brighter systems.


\subsection{Effects of Irradiation}
\label{subsec:irradiation}

\cite{Deloye2007} showed that irradiation can slow the cooling of the donor. First, the luminosity of the donor is set by the opacity at the surface radiative-convective boundary \citep[e.g.,][]{Arras2006}. By keeping the surface at higher temperatures, where the opacities tend to be higher, irradiation decreases the intrinsic luminosity of the donor and slows its cooling. 
Second, the cooling of the donor is significantly faster once it becomes fully convective. Irradiation keeps the surface boundary at a higher temperature, and hence makes the temperature gradient less steep, which tends to inhibit convection \citep[this is also seen in models of strongly irradiated black-widow pulsar companions; e.g., ][]{ConradBurton2023}. While \cite{Deloye2007} focused on the effects of irradiation on the mass-radius relation of the donor, here we explore how it in turn changes the orbital evolution and manifests in the luminosity evolution of the accretor. 

The left panel of Figure~\ref{fig:irradiation} shows the effects of irradiation on the donor's mass-radius evolution, for a $\iniSc = 3.7\,\kB \NA$ donor. 
As expected, irradiation slows the cooling of the donor, which prolongs the adiabatic phase at short $\Porb$ and allows the donor to remain puffy at longer $\Porb$, in better agreement with observations. Nevertheless, there still comes a time when $ \tauth < \taumdot$, and the donor still cools to the cold WD relation at $\Porb \approx 60$~min. This is even with an optimistic efficiency of $\eta = 0.5$, and the optimistic assumption that the irradiating flux is given by the accretion luminosity set by the secular $\mdot$, instead of the stellar luminosity of the accretor. We thus conclude that while irradiation is important in slowing the cooling of the donor, in better agreement with observations of the mass-radius relation of the donor, it is only part of the solution. 

The right panel of Figure~\ref{fig:irradiation} shows the accretor $\Teff$ evolution with an irradiated donor. With irradiation, the donor cools more slowly and is more massive at a given $\Porb$ (see the left panel), which allows for a faster orbital evolution. In turn, at long $\Porb$, the accretor is younger and hence hotter. The difference between our model without irradiation (blue solid line), and one with $\eta = 0.5$ (green dotted line), can reach $\approx 1000$~K at $\Porb \approx 50$~min. 
This is an improvement in the right direction, but as we have shown in Section~\ref{subsec:general features}, the observed systems are hotter than predicted even with a donor evolving adiabatically. 

Our irradiation treatment is a simplified approach that ignores the detailed deposition of the irradiating flux, non-grey effects, redistribution of heat across the donor surface, etc., which we defer to future studies \citep[e.g.,][]{Lothringer2020}. One interesting observable that remains to be tested, is the SED of the irradiated donor in the infrared. Our irradiated models generally predict a donor $\Teff \approx 1000-2000$~K at $\Porb = 50-60$~min, which, if modeled as a blackbody, appears broadly consistent with the level of infrared excess in AM~CVn systems (e.g., \citealt{Green2020}, \citetalias{vanRoestel2022}, \citealt{Macrie2024}). However, recent James Webb Space Telescope (JWST) observations of three eclipsing systems show that most of the infrared excess originates from the accretion disk \citep[][]{ElBadry2026}. 
Constraining the $\Teff$ of the irradiated donor will require radiative transfer modeling of the cold, dense He atmosphere. 

In summary, while irradiation can slow the cooling of the donor and allow for faster orbital evolution, additional solutions are required, such as extra AML.

\subsection{Varying opacity}
\label{subsec:opacity}

As we have previously shown, the surface opacity of the donor is important in determining its intrinsic luminosity and thermal evolution. Our base models adopt the post-H burning abundances of a $1.5\,\msun$, solar-metallicity ($Z=Z_{\odot}=0.0142$) star. 

We checked the sensitivity of our models to the adopted opacities, by varying the underlying abundance patterns and metallicities. 
In particular, the zero-age main sequence mass (ZAMS) of the donor progenitor affects products of the CNO bi-cycle \citep[e.g.,][]{Nelemans2010}, and hence the molecular opacities. 
We therefore fixed the metallicity ($Z=Z_{\odot}$) and obtained additional \aesopus~low-temperature opacity tables, computed with the post-H burning abundances of $1\,\msun$ and $1.3\,\msun$ stars. 
In addition, we obtained opacity tables for a $1.5\,\msun$, low-metallicity ($Z=0.1Z_{\odot}$) star. 
We recalculated the binary evolution of our $\iniSc=3.7\,\kB\,\NA$ donor, using these new opacity tables. 
At low temperatures, the opacities with a $1\,\msun, Z_{\odot}$ progenitor are higher than our fiducial, due to a higher oxygen abundance which affects the molecules formed, whereas those with a $1.5\,\msun, 0.1 Z_{\odot}$ progenitor are lower than our fiducial. We discuss these opacities in detail in Appendix~\ref{sec:appendix opacity table}. 

\begin{figure*}[t!]
\centering
\fig{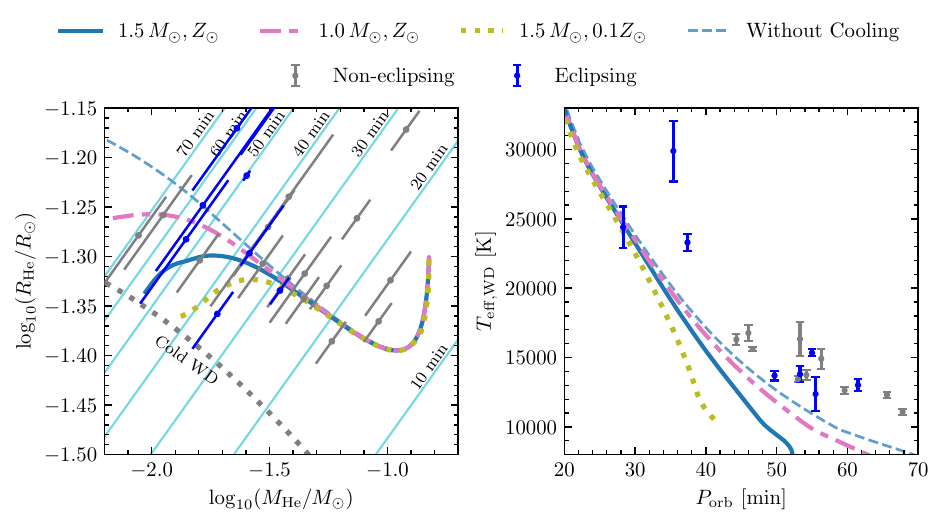}{ \textwidth }{}
\vspace{-25 pt}
\caption{ 
Effects of varying opacity on the donor mass-radius relation (left) and accretor effective temperature (right) of our models with a $\iniSc = 3.7\,\kB \NA$ donor. 
Blue solid line uses the post-H burning abundances of a $1.5\,\msun$, solar metallicity star, pink dashed line uses that of a $1.0\,\msun$, solar metallicity star, and yellow dotted line uses that of a $1.5\,\msun$, one-tenth solar metallicity star. 
Blue dotted line shows the limit where the donor evolves adiabatically (i.e. without cooling). 
Data and other annotations same as Fig.~\ref{fig:irradiation}. 
\label{fig:opacity_mass_evolution}}
\end{figure*}

The resulting mass-radius evolution of the donor, and the evolution of the accretor $\Teff$, are shown in Figure~\ref{fig:opacity_mass_evolution}. 
With a low ZAMS mass such as the $1\,\msun$, the opacities are higher. This leads to a slower cooling of the donor, as seen in its mass-radius evolution, and in turn a faster orbital evolution. The accretor is then younger and hotter at a fixed $\Porb$. 

Conversely, the lower opacities with $0.1Z_{\odot}$ cause an earlier onset of the cooling of the donor, already at $\Porb  = 40$~min, and yield a cooler accretor. In comparison with our fiducial models presented in Section~\ref{subsec:general features}, the low metallicity models struggle even more with explaining the AM~CVn population. 
This makes the case for extra AML even stronger because, 
while abundance measurements are scarce, at least two are thought to be metal-poor from the non-detection of Si, namely the 46~min GP~Com \citep[][]{Marsh1991} and the 65~min V396~Hya \citep[][]{Roelofs2005,Nagel2009}. 

The cooling of the donor, and its susceptibility to the opacity variations presented here, should in principle be germane to all three formation channels \citep[see also][]{Deloye2007}. 
We expect that the opacities presented here to apply well to the evolved~CV channel after period bounce when the donor surface becomes H-deficient. 
For the He~star channel, the occurrence of core He burning changes the C/O ratio of the donor, which is particularly important for molecular opacities. This should be accounted for in future studies. 


\section{Evolution with magnetic braking}
\label{sec:MB}

We have shown in Section~\ref{sec:GWR only} that at long $\Porb$, the accretor behaves like a cooling WD and its temperature reflects the orbital evolution set by the donor. Nevertheless, even in the optimistic case where the donor evolves adiabatically, observed AM~CVn binaries appear brighter than predicted. Accounting for realistic cooling of the donor, the problem worsens: the discrepancy in accretor luminosity is even larger, and the inflated radii of AM~CVn donors cannot be reproduced. 

Both of these problems can be circumvented if there is extra AML: (1) the faster orbital evolution allows for a younger, and hence brighter, accretor at a given $\Porb$ (see also \citealt{Maccarone2024}), and (2) the mass transfer timescale may become shorter than the donor thermal timescale, such that there is no time for the donor to cool (see also \citealt{Belloni2023}). 

A natural explanation for the extra AML is MB of the donor, where the angular momentum of the donor is efficiently extracted by a magnetized wind corotating out to large radii\footnote{\label{footnote5}This requires small amounts of mass loss or non-conservative mass transfer, which we do not explore here. In principle, non-conservative mass transfer by itself leads to AML. However, we found that for reasonable mass loss rates from either stellar components or the L2 Lagrange point, the AML is not large enough to reproduce observations.}. Historically, the form of MB was empirically inferred from the rotation rates of single, mid-F- and later-type stars, which spin down with age as $\Prot \propto t^{1/2}$, implying a torque $\Jdotmb \propto \Prot^{-3}$ \citep[][]{Kraft1967,Skumanich1972}. 
Nevertheless, the Skumanich law does not appear to be universal. 
For example, MB in these stars beyond the solar age appears to weaken compared to standard models \citep[e.g.,][]{vanSaders2016,vanSaders2019,Metcalfe2025}. 
Fast rotating stars appear to experience ``saturated'' MB, where the torque scales more gently with $\Prot^{-1}$ instead of the $\Prot^{-3}$ \citep[e.g.,][]{Sills2000,ElBadry2022}, below a critical Rossby number $\Rocrit \approx 0.1$, where the Rossby number is given by $\Ro = \Prot / \tauconv$, and $\tauconv$ is the convective turnover time. 
Below this $\Rocrit$, stellar activity also appears to saturate to a constant value, which hints at changes in the dynamo-generated magnetic fields that also drive the AML \citep[e.g.,][]{Noyes1984,Reiners2009,Wright2011}. 
While there have been attempts at deriving a first-principles theory from analytics \citep[e.g.,][]{Weber1967,Mestel1987,Kawaler1988} and magnetohydrodynamical simulations \citep[e.g.,][]{Matt2012,Reville2015}, there is no generally agreed-upon prescription. 

\begin{deluxetable*}{cccc}
\tablenum{1}
\label{tab:torque_estimates}
\tablecaption{Torque Scalings and Estimates}
\tablehead{
\colhead{AML Mechanism} & 
\colhead{Explicit $\Jdot$ Scaling} & 
\colhead{Estimated $\Jdot$ Scaling\tablenotemark{a}} & 
\colhead{Estimated $\Jdotmb/\Jdotgr$\tablenotemark{b}} 
}
\tablewidth{100pt}
\startdata
GWR & $\propto \Porb^{-7/3} \mhe^{2} \mwd^{2} (\mhe + \mwd)^{-2/3} $ & $\appropto \Porb^{-5}$ & - \\
\rvj & $\propto \Porb^{-3} \mhe \Rhe^{3} $ & $\appropto \Porb^{-3} $ & $\approx 0.5$ \\
\sills & $\propto \Porb^{-1} \mhe^{-1/2} \Rhe^{1/2} \tauconv^{-2} $ & $\gtrpropto \Porb^{-4} $ & $\approx 100$ \\
\matt & $\propto \Porb^{-1} \mhe^{1/2} \Rhe^{3.1} $ & $\appropto \Porb^{-1} $ & $\approx 0.03$ \\
\carb & $\propto \Porb^{-11/3} \mhe^{-2/3} \Rhe^{16/3} \tauconv^{8/3} \mdotw^{-1.3} $ & $\appropto \Porb^{1} \mdotw^{-1/3} $ & $\approx 1000$ \\
\enddata
\tablecomments{
\tablenotemark{a}{These estimations, derived analytically assuming $\xi = -0.2$, agree with models generally. Our $\mesa$ models show $\JdotSills \appropto \Porb^{-(5-6)}$, and $\JdotCARB \appropto \Porb^{0.5}$. }
\tablenotemark{b}{Estimated at $\Porb = 60$~min. } 
} 
\end{deluxetable*}

MB in binaries remains poorly constrained. 
In CVs, the scarcity of systems in the ``period gap'' between 2 and 3 hrs, is explained by substantial weakening of MB when the donor becomes fully convective \citep[e.g.,][]{Rappaport1983,Howell2001}. 
Nevertheless, there is no obvious sign of a drop in MB strength near the fully convective boundary in single stars \citep[][]{Lu2024}, wide WD+MS binaries \citep[][]{Chiti2024}, and low-mass detached close binaries \citep[][]{ElBadry2022}; if anything, the first two studies find a stronger AML in fully convective stars, and a change in dynamo behavior and/or magnetic field topology near the fully convective boundary has been suggested as a possible explanation. 
On the other hand, some recent binary studies have favored a MB prescription that is ``saturated'' (scales as $\Porb^{-1}$), ``boosted'' by a factor $\approx 20 - 100$ compared to single star prescriptions, and ``disrupted'' when fully convective \citep[e.g.,][]{Belloni2024,Blomberg2024}. Irradiation of the donor may play an important role in enhancing MB \citep[][]{Dodon2026}.

Given all the uncertainties of MB, we experimented with four MB prescriptions that are popular in the literature, and evolved AM~CVn binaries including AML from MB. 
These are, \citealt{Rappaport1983} (hereafter \citetalias{Rappaport1983}), \citealt{Matt2015} (hereafter \citetalias{Matt2015}), \citealt{Sills2000} (hereafter \citetalias{Sills2000}), and \citealt{Van2019} (hereafter \citetalias{Van2019}). 
We summarize in Table~\ref{tab:torque_estimates} their explicit $\Porb$ scaling, and our estimated $\Porb$ scaling assuming $\xi = -0.2$. We also show our estimated ratio between $\Jdotmb$ and $\Jdotgr$ at $\Porb = 60$~min. These estimates agree well with our \mesa~models. We detail our semi-analytical estimates in Appendix~\ref{sec:appendix MB}. 

We are well aware that most of the MB prescriptions are calibrated for solar-type stars, and do not necessarily apply well to AM~CVn binaries. 
Our goal is to \textit{empirically} derive the extra AML required to explain the evolution of AM~CVn binaries, and test whether any of the pre-existing prescriptions can be generalized. 

All simulations take a $\iniSc = 3.7 \, \NA \kB$ donor, and a $\iniMwd = 0.7\,\msun$~accretor, without irradiation on the donor. The donor entropy and accretor mass were chosen to reproduce the observed donor mass-radius relation, and the final accretor mass $\approx 0.8 - 0.85\,\msun$. 
For numerical stability, we cap $\Jdotmb$ at a maximal amplitude of $5 \times 10^{35}\,\Jdotunit$; models that require this treatment (\citetalias{Sills2000}~MB) cannot reproduce observations anyways. 

For prescriptions that depend on the convective turnover timescale $\tauconv$, namely \citetalias{Sills2000} \& \citetalias{Van2019}~MB, we calculated $\tauconv = H / v_{\rm conv}$, at $0.5 H$ above the base of the convection zone, where $H$ is the local pressure scale height and $v_{\rm conv}$ is the local convective velocity, following \cite{Gossage2025}. 
Under rapid mass transfer, especially near the period minimum, we found that spurious convection zones can appear in the donor, leading to jumps in $\tauconv$. For numerical stability, we imposed a time-smoothing scheme, by requiring that $\tauconv$ change less than 1\% between time steps (for \citetalias{Van2019}~MB, which is sensitive to changes in $\tauconv$ as $\JdotCARB\propto \tauconv^{8/3}$, this is tightened to 0.1\%). The resulting $\tauconv$ still accurately tracks the long-term behavior of the unsmoothed version, but shows much stabler behavior between time steps. 

We find that $\tauconv$ ranges from $\sim 1$~d at period minimum to $\sim 10^{3}$~d at long $\Porb$, yielding a Rossby number $Ro \sim 10^{-2} - 10^{-4} \ll 1$. As a result, the \citetalias{Matt2015} and \citetalias{Sills2000} MB formalisms are always in the saturated regime with an explicit $\Jdotmb \propto \Porb^{-1}$ scaling. 

\begin{figure*}[t!]
\centering
\fig{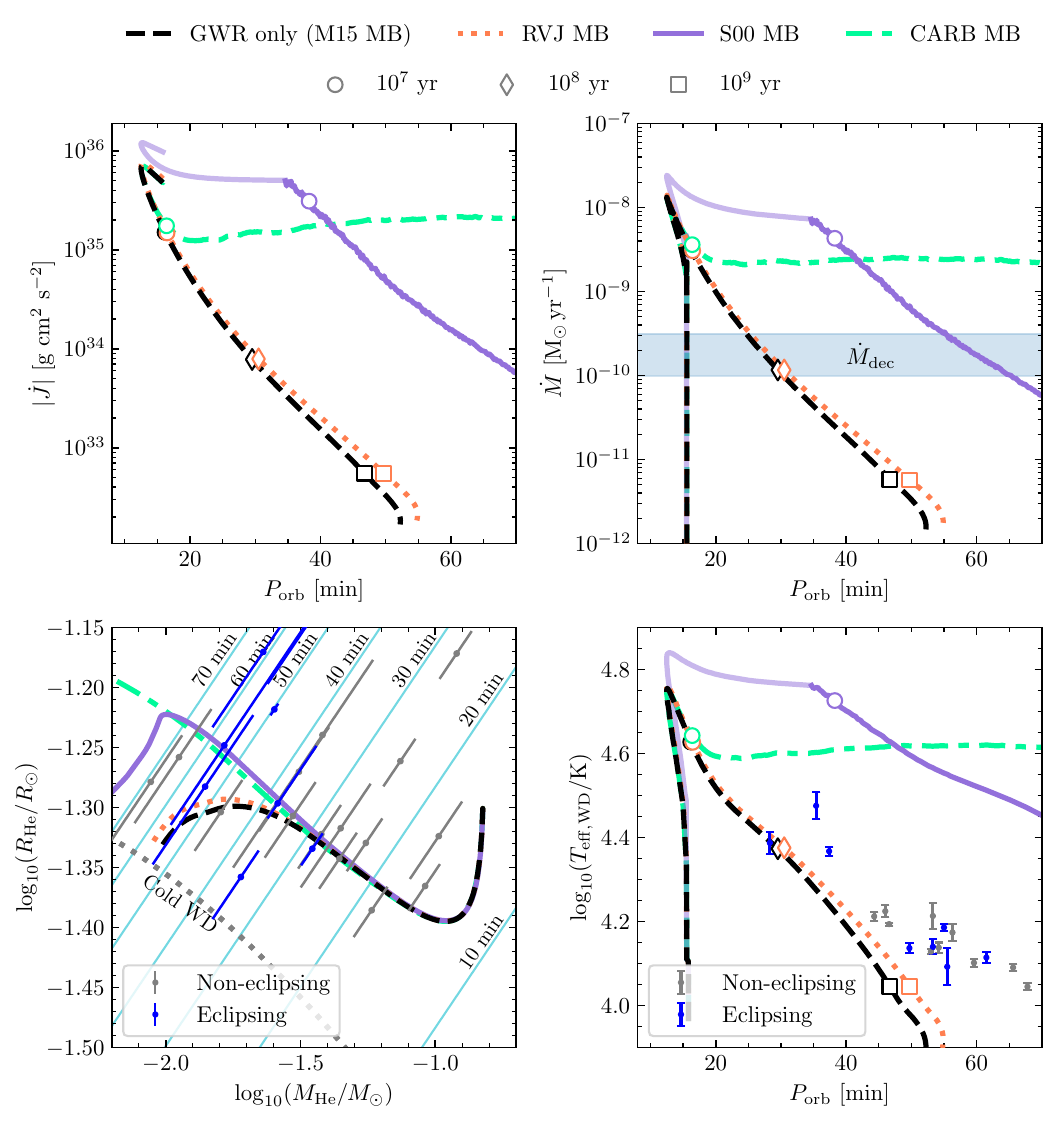}{ 0.8 \textwidth }{}
\vspace{-15 pt}
\caption{ 
Evolutionary tracks with various MB prescriptions. 
The black dashed line shows a model with only GWR. The other models include various MB AML in addition to GWR, namely RVJ~MB (coral dotted), S00~MB (purple solid), and CARB~MB (green dot-dashed). Including M15~MB results in a track identical to the black, which is not shown. The ligher shade in the S00 MB model shows when we cap $|\Jdotmb|$ to a maximum value of $ 5 \times 10^{35} \, \Jdotunit$. 
The markers show the age since the start of mass transfer, at $10^{7}$ (circles), $10^{8}$ (diamonds), and $10^{9}$~yrs (squares). 
\textit{Top left}: total torque. 
\textit{Top right}: mass transfer rate. Blue shaded region shows the decoupling mass transfer rate, below which the accretor cools. 
\textit{Bottom left}: donor mass-radius relation. 
\textit{Bottom right}: accretor effective temperature. 
Data and other annotations follow Fig.~\ref{fig:MR_and_Mdot_GW} and \ref{fig:Teff_and_Mg_GW}. 
\label{fig:MB_evolution}}
\end{figure*}


\subsection{Evolution with Original MB Strengths}
\label{subsec:original MB}

Figure~\ref{fig:MB_evolution} shows evolutionary tracks with various MB prescriptions. The top left panel shows the total torque, $\Jdot = \Jdotgr + \Jdotmb$, top right panel shows the mass transfer rate $\mdot$, bottom left panel shows the donor mass-radius relation, and the bottom right panel shows the accretor effective temperature. We compare the latter two to observations, as discussed in Sections~\ref{subsubsec:GWR donor evolution} \& \ref{subsubsec:GWR comparison to observations}. 
For comparison, we show the GWR-only track (black dashed line), as presented in Section~\ref{subsec:general features}. 
We do not show a model that includes \citetalias{Matt2015}~MB, since its torque is much weaker compared to GWR ($\JdotMatt \approx 0.03 \times \Jdotgr$ at $\Porb\approx 60$~min) and the resulting evolutionary model is virtually identical to the GWR-only model. 
Despite being a saturated MB formalism like \citetalias{Sills2000}~MB, \citetalias{Matt2015}~MB is much weaker due to its dependence on stellar structure, particularly its steeper $\propto \Rhe^{3.1}$ scaling (compared to $\propto \Rhe^{0.5}$ in the former). 

\bigskip

\textbf{\citetalias{Rappaport1983}~MB}: At short $\Porb$, \citetalias{Rappaport1983}~MB only changes the total torque by $\lesssim 10\%$, but its effects increase to $\approx 30\%$ at longer $\Porb$, since it scales more shallowly with $\Porb$ than $\Jdotgr$: 
$\Jdotgr \appropto \Porb^{-5}$, whereas $\JdotRVJ \appropto \Porb^{-3}$. 
The orbital evolution with \citetalias{Rappaport1983}~MB is slightly faster, as shown by the markers labeling the age since the start of mass transfer (e.g. squares show $10^{9}$~yrs). 
The cooling of the donor is delayed slightly, but still occurs at $\Porb \lesssim 60$~min, and so cannot reproduce the inflated donor radii at the longest $\Porb$ (see bottom left panel). 
The faster orbital evolution also yields a slightly higher accretor temperature, in mildly better agreement with observations (see bottom right panel). 

\bigskip

\textbf{\citetalias{Sills2000}~MB}: 
As we estimated in Section~\ref{subsubsec:S00}, $\JdotSills$ is orders-of-magnitude larger than $\Jdotgr$. As a result, the model evolves past $\Porb = 70$~min within $\approx 5 \times 10^{7}$~yrs. At $\Porb \lesssim 35$~min, $\JdotSills$ exceeds our cap of $5 \times 10^{35}\,\Jdotunit$ (shown with lighter shade), imposed for numerical stability. For most of the range shown, the mass transfer rate is above that for decoupling, $\mdot \gtrsim \mdotdec$ (see top right panel), meaning that the accretor effective temperature, $\Teffwd$, is set by compressional heating and not WD cooling (\citetalias{Bildsten2006}). Consequently, $\Teffwd$ stays hot above 25,000~K, way hotter than any of the observations at $\Porb \gtrsim 40$~min (see bottom right panel). 

However, despite its shortcomings, this case does show that rapid orbital evolution can effectively prevent the donor from cooling: since the orbital evolution timescale is much faster than the donor thermal timescale, the donor evolves adiabatically and remains inflated out to $\Porb = 70$~min, in agreement with observations (see bottom left panel). 

Finally, we note that despite the explicit $\Porb^{-1}$ scaling of $\JdotSills$, its decline with $\Porb$ is a lot steeper. 
We found that $\tauconv \appropto \Porb^{3} $ for this model, which leads to $\Jdot \propto \tauconv^{-2} \appropto \Porb^{-6}$ and dominates the scaling of $\JdotSills$ with $\Porb$. 
Since $\JdotSills$ has a steep drop-off with $\Porb$, it weakens significantly at long $\Porb$. Despite the nearly adiabatic evolution in the range shown in Figure~\ref{fig:MB_evolution} (up to $\Porb = 70$~min), the donor becomes fully convective and cools at $\Porb \approx 75$~min. 

\bigskip

\textbf{\citetalias{Van2019}~MB}: 
This is another example where $\Jdotmb$ overpowers $\Jdotgr$ and leads to a rapid orbital evolution. 
Near the period minimum, $|\JdotCARB| \lesssim |\Jdotgr|$, so initially, the system evolves similarly to without \citetalias{Van2019}~MB. 
As the system evolves to longer $\Porb$, one might naively expect a declining $\JdotCARB$ due to its explicit $\propto \Porb^{-11/3}$ scaling. 
This is not the case, because (1) $\JdotCARB$ also scales as $\tauconv^{8/3}$, and $\tauconv$ generally increases with $\Porb$ (see the discussion in Section~\ref{subsec:theoretical expectations})\footnote{For this model, we found a scaling $\tauconv \propto \Porb$, shallower than the other models. Due to its much more rapid orbital evolution, the base of the convective envelope recedes to lower temperatures and densities with $\Porb$, yielding a more shallowly increasing $H$ and hence $\tauconv$.}, and (2) $\JdotCARB$ has a scaling $\propto \mhe^{-2/3} \Rhe^{16/3} \appropto \Porb^{2.17}$. The combination of these cancel out the explicit $\propto \Porb^{-11/3}$ scaling, and lead to a torque roughly constant with $\Porb$. 
The ``convective boosting'' maintains a very strong MB as the orbit widens, while $\Jdotgr \appropto \Porb^{-5}$ declines sharply with $\Porb$, leading to $|\JdotCARB| \gg |\Jdotgr|$. 
The system rapidly evolves beyond $\Porb = 70$~min in $\approx 3 \times 10^{7}$~yrs. 

Due to the rapid orbital evolution, $\mdot$ remains at $\approx 10^{-9}\,\msunyr$, above the decoupling regime, resulting in a hot accretor effective temperature $\approx 4 \times 10^{4}$~K, that remains roughly constant with $\Porb$ (see bottom right panel). Not only is the model $\Teffwd$ much hotter than observed, but the model also does not show the decline with $\Porb$ that is seen in the data. 

On the bright side, the rapid orbital evolution due to \citetalias{Van2019}~MB allows for adiabatic evolution of the donor all the way out to long $\Porb$ (see bottom left panel). The donor remains inflated, in agreement with observations. In fact, this was one of the motivations for \cite{Belloni2023} to keep \citetalias{Van2019}~MB on after period minimum in their evolved~CV models (see their Figure~7).

\bigskip

\textbf{Summary:} None of the MB prescriptions in their original forms can reproduce the observations: 
\citetalias{Sills2000} and \citetalias{Van2019}~MB produce too rapid orbital evolution, such that the accretor is much hotter than observed, but the inflated radii of the donor are reproduced; the orbital evolution due to \citetalias{Rappaport1983} and \citetalias{Matt2015}~MB is insufficient to prevent the donor from cooling at long $\Porb$, and the accretor is cooler than observed. In order to reproduce the observations, a MB prescription with strength intermediate between these two classes is required, which we will show next. 


\subsection{Varying MB Strength}
\label{subsec:varying MB strength}

\begin{figure*}[t!]
\centering
\fig{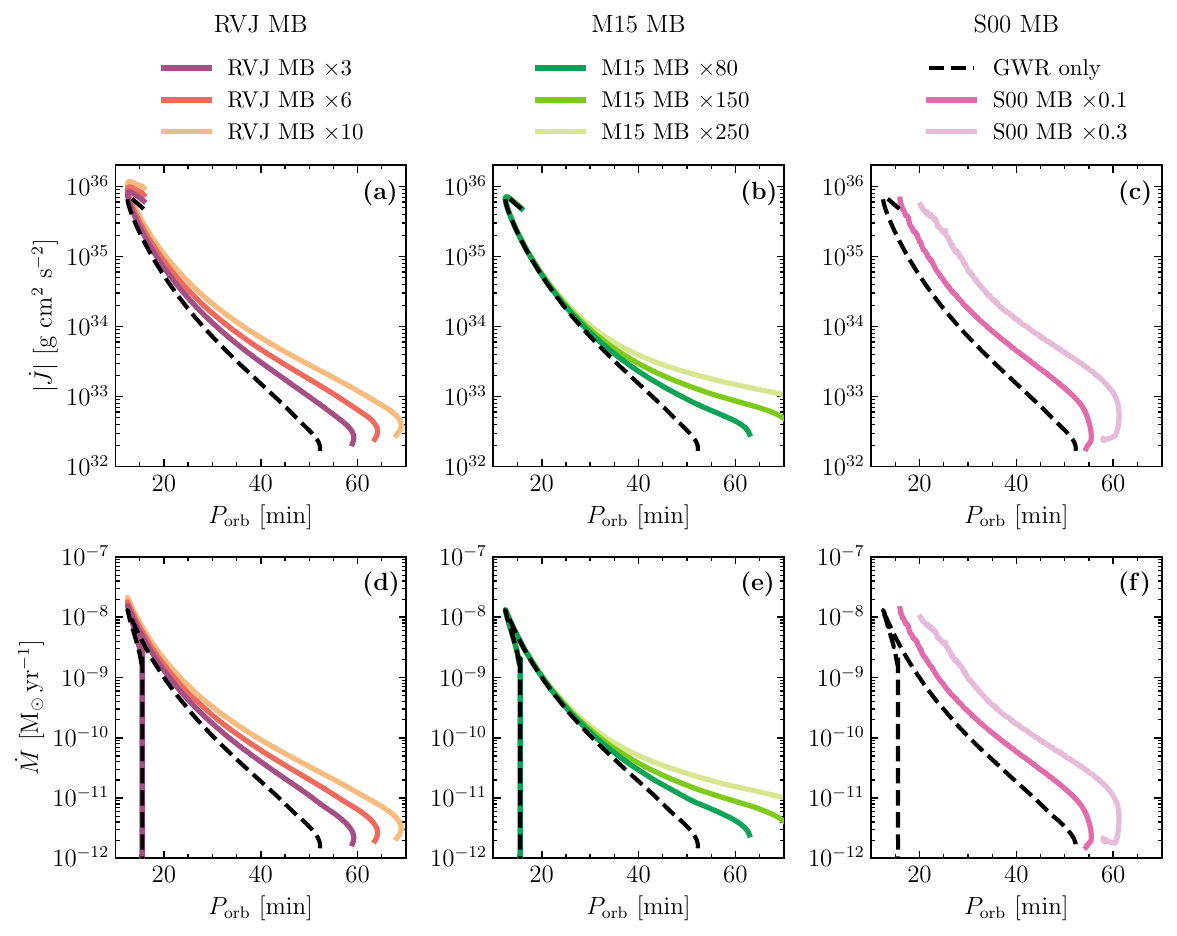}{ \textwidth }{}
\vspace{-25 pt}
\caption{ 
Total torque (top row) and mass transfer rate (bottom row), for various MB prescriptions scaled by different factors. We show a model where the only torque is due to GWR in black dashed line. 
Left column: \rvj scaled by factors $3-10$. 
Middle column: \matt scaled by factors $80-250$. 
Right column: \sills scaled by factors $0.1 - 0.3$. We cut out parts of the evolution where $-\Jdotmb$ exceeds our cap $5 \times 10^{35}\,\Jdotunit$. 
\label{fig:MB_evolution2}}
\end{figure*}

\begin{figure*}[t!]
\centering
\fig{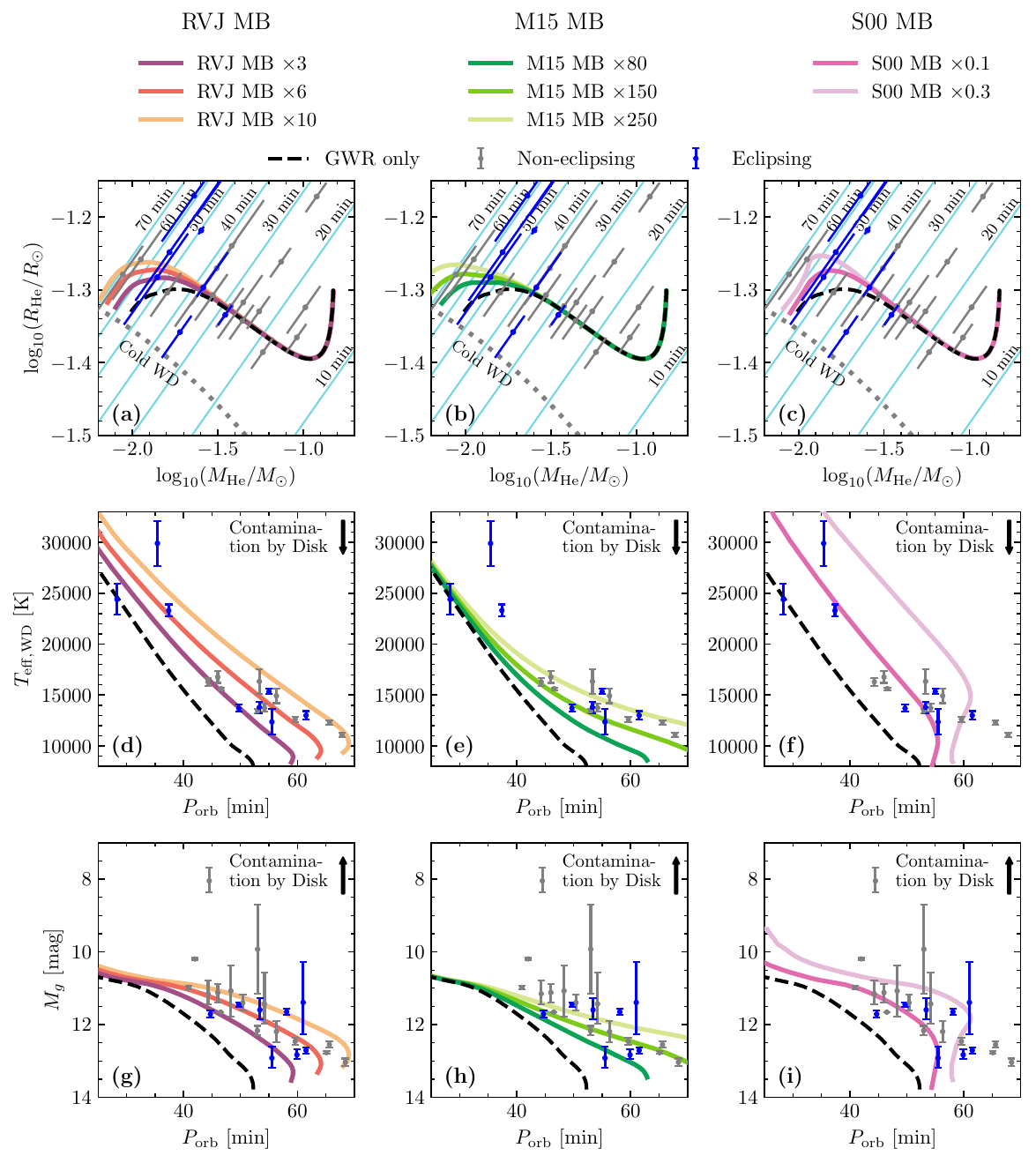}{ \textwidth }{}
\vspace{-25 pt}
\caption{ 
Donor mass-radius relation (top row), accretor effective temperature (middle row), and $g$-band absolute magnitude (bottom row), for various MB prescriptions scaled by different factors. 
The arrows show that contamination by the accretion disk tends to lower the $\Teffwd$ inferred from the SED (middle row) and make the system brighter (bottom row), especially at shorter $\Porb$. 
Left column: \rvj scaled by factors $3-10$. 
Middle column: \matt scaled by factors $80-250$. 
Right column: \sills scaled by factors $0.1 - 0.3$. 
Data and other annotations follow Fig.~\ref{fig:MR_and_Mdot_GW} and \ref{fig:Teff_and_Mg_GW}. 
\label{fig:MB_evolution2_observables}}
\end{figure*}

Here we show models scaling the normalization of the MB prescriptions, in order to empirically determine the amount of MB needed to reproduce observations. Similar exercises were taken in other binary scenarios, namely detached close low-mass MS binaries with WD companions \citep[][]{Belloni2024}, and with subdwarf B star companions \citep[][]{Blomberg2024}. 

We show the resulting total torque and mass transfer rate in Figure~\ref{fig:MB_evolution2}. 
We show models where \rvj is increased by a factor $3-10$ (left column), \matt is increased by a factor of $100 - 300$ (middle column), and \sills is $0.1 - 0.3$ weaker (right column). 
We did not attempt a normalization experiment with \citetalias{Van2019}~MB, because we would still expect a flat evolution in the accretor luminosity with $\Porb$, which is in strong tension with observations. 

We compare the corresponding donor mass-radius relation, accretor effective temperature $\Teffwd$, and accretor $g$-band absolute magnitude $M_{g}$, to observations, in Figure~\ref{fig:MB_evolution2_observables}. For the comparison to observations of the accretor, we remind the reader that \textit{contamination by the accretion disk increases towards shorter $\Porb$}, and so (1) blackbody fits to the SED will likely yield lower $\Teffwd$, since the accretion disk contaminates more strongly at longer wavelengths (e.g., \citetalias{vanRoestel2022}), and (2) the system may be more luminous than the accretor photosphere alone. 
We indicate in arrows the effects of disk contamination on the accretor temperature (middle row) and the $g$-band magnitude of the system (bottom row). 
We strongly advise caution when comparing the data at $\Porb \lesssim 50$~min to models (see also the discussion in Section~\ref{subsubsec:GWR comparison to observations}). 
The only exceptions are YZ~LMi, ZTF~J0407-0007, ZTF~J2252-0519 at $\Porb = 28 - 37$~min, which have $\Teffwd \approx $~23,000$ - $30,000~K measured from HST UV spectroscopy \citep[][]{Yu2026,Yu2026b}.

\bigskip

\textbf{\citetalias{Rappaport1983}~MB}: 
We find that boosting \rvj by a factor somewhere between 3 and 10 can yield decent matches to observations. 
Near period minimum, these models show $\approx 20 - 50\%$ deviation from $\Jdotgr$ but become increasingly dominant at longer $\Porb$, due to its shallower scaling with $\Porb$ than $\Jdotgr$. At $\Porb = 60$~min, $\Jdotmb$ is about $1.5 - 5$ times that of GWR (see Figure~\ref{fig:MB_evolution2}, panel a). 

With the moderate amount of extra AML, the cooling of the donor is delayed to $\Porb \approx 60 - 70$~min, and thus reproduces the inflated radii of systems across most of the observed period range (see Figure~\ref{fig:MB_evolution2_observables}, panel a). 
While models with scaling factors $\lesssim 6$ do not seem to extend out to the longest $\Porb$, this can be remediated by considering irradiation of the donor (see Section~\ref{subsec:irradiation}). 

Ignoring the HST-observed systems at $\Porb = 28 - 37$~min for the moment, these models appear to show a steeper trend of $\Teffwd$ with $\Porb$, than observationally inferred (Figure~\ref{fig:MB_evolution2_observables}, panel d). Nevertheless, contamination by the accretion disk is more severe at shorter $\Porb$, and since its effect is to decrease the inferred $\Teffwd$, we expect the true $\Teffwd-\Porb$ relation to be steeper than inferred. If we now consider the HST-observed systems, the scaled \rvj models appears broadly consistent with the overall trend. 

The scaled \rvj models show excellent agreement with the observed $M_{g}-\Porb$ relation (Figure~\ref{fig:MB_evolution2_observables}, panel g). However, the expected stronger disk contamination at shorter $\Porb$ means that the true $M_{g}-\Porb$ relation is shallower. This likely means that the underlying extra AML scales more shallowly with $\Porb$, than \rvj ($\propto \Porb^{-3}$).

\bigskip

\textbf{\citetalias{Matt2015}~MB}: Models that scale \matt by a factor $\approx 80 - 250$ also show general agreement with observations. Its shallower $\Porb^{-1}$ scaling means that it does not lead to much correction to $\Jdotgr$ near the period minimum ($\lesssim 2\%$), while becoming increasingly important at longer $\Porb$ (see Figure~\ref{fig:MB_evolution2}, panel b). We similarly infer $\Jdotmb \approx (3 - 7) \times \Jdotgr$ at $\Porb = 60$~min . 

Compared to the scaled \rvj models, the scaled \matt models delay cooling of the donor out to slightly longer $\Porb$ (Figure~\ref{fig:MB_evolution2_observables}, panel b). This is due to its shallower scaling with $\Porb$ compared to \citetalias{Rappaport1983}~MB, which allows it to stay relatively strong at long $\Porb$. 

The models appear to agree well with the inferred $\Teffwd$ from \cite{Macrie2024} (Figure~\ref{fig:MB_evolution2_observables}, panel e). Bearing in mind that the shorter $\Porb$ systems likely have underestimated values, this means the true torque may scale slightly more steeply than $\Porb^{-1}$. The models underestimate the inferred $\Teffwd$ of two out of three $\Porb = 28 - 37$~min HST-observed systems \citep[][]{Yu2026,Yu2026b}. However, with the small sample size, it is unclear whether the slight offset from the models is due to a slightly more massive accretor, or a higher-entropy donor, than assumed here; for instance, \cite{Yu2026} showed that ZTF~J2252-0519 can be explained by the combination of a $\iniMwd = 0.90\,\msun$ accretor and a $\iniSc = 3.9\,\NA\kB$ donor without invoking extra AML. Given all this, we conclude that the scaled \matt models are consistent with the observed $\Teffwd$. 

The models also perform well in predicting the $g$-band magnitudes (Figure~\ref{fig:MB_evolution2_observables}, panel h). The models fit the data well at long $\Porb$, and although the models are slightly underluminous compared to data at short $\Porb$, this is qualitatively what we expect for contamination by the accretion disk. 

\bigskip

\textbf{\citetalias{Sills2000}~MB}: 
We tested models where \sills is scaled by factors $0.1 - 0.3$. As with the original prescription, these models predict strong MB at short $\Porb$, exceeding our imposed cap of $5 \times 10^{35}\,\Jdotunit$, which we cut off from Figure~\ref{fig:MB_evolution2} (right column). The shape of $\JdotSills$ is similar to that of $\Jdotgr$, due to a similar scaling with $\Porb$ ($\appropto \Porb^{-5}$). 

Although \sills can significantly accelerate orbital evolution at short $\Porb$, it weakens significantly at longer $\Porb$ due to its steep scaling with $\Porb$. 
Consequently, the donor cools and becomes fully convective at $\Porb \approx 55 - 60$~min. 
So while \sills can reproduce the inflated donor radii at shorter $\Porb$ by virtue of its strong torque, it fails to do so at the longest $\Porb$. 

Correspondingly, the model with scaling factor 0.1 is generally consistent with the observed $\Teffwd$ and $M_{g}$ at $\Porb \lesssim 50$~min, including ZTF~J2252-0519. The one with scaling factor 0.3 slightly overpredicts both of these quantities, but fits well the spectroscopically derived $\Teffwd$ of ZTF~J0407-0007. Due to the early cooling of the donor, neither models can explain the longest-period systems.

The above considerations suggest that the scaling of \sills with $\Porb$ may be too steep: since the torque weakens significantly at long $\Porb$, the donor rapidly cools, and the model fails to explain the systems at $\gtrsim 60$~min. However, models with scaling factors $0.1 - 0.3$ fit well ZTF~J0407-0007 and J2252-0519 at $\Porb = 35$ and $37$~min. 

\bigskip

\textbf{Summary}: 
Overall, the observations constrain the amplitude and the effective evolution with $\Porb$ of the extra AML. We find that \sills declines too steeply with $\Porb$, \carb remains too strong at long $\Porb$, but with the appropriate normalization, \rvj ($\appropto \Porb^{-3}$) and \matt ($\appropto \Porb^{-1}$) can reproduce the observations. \textit{The extra AML becomes increasingly important relative to GWR at long $\Porb$}: it is subdominant compared to GWR at short $\Porb$ (few to tens of \%), but it gradually grows to about $3 - 7$ times that of GWR, at $\Porb = 60$~min. 
Whether scaled \citetalias{Rappaport1983} or \citetalias{Matt2015}~MB, the required MB yields $\Jdotmb \approx -5 \times 10^{32}\,\Jdotunit$ at $\Porb=60$~min. We cannot confidently favor MB of the classical flavor (e.g., \citetalias{Rappaport1983}~MB) or the saturated flavor (e.g., \citetalias{Matt2015}~MB) over the other, due to the poorly constrained contamination by the accretion disk at short $\Porb$, and the lack of spectral modeling of an AM~CVn accretor except for 3 systems. 

However, the $\Porb \gtrsim 50$~min eclipsing systems from \citetalias{vanRoestel2022} show $\lesssim10\%$ contamination in the optical by the accretion disk/hotspot, corresponding to a magnitude difference of only $\approx 0.1$~mag, much smaller than the scatter among systems with similar $\Porb$. 
Comparison between blackbody and DB atmosphere fits shows a difference in $\Teffwd$ by $\approx 1000$~K. 
We therefore believe that the range of our MB models appropriately captures the required normalization of $\Jdotmb$. 
Furthermore, the extra torque is unlikely steeper than that effectively given by \sills ($\appropto \Porb^{-5}$), since the torque would weaken significantly at long $\Porb$ and the donor would rapidly cool. 
The $\Porb = 28 - 37$~min systems from the HST campaign by \cite{Yu2026,Yu2026b} are useful anchors for probing the $\Porb$-dependence, and tentatively favor scaled \citetalias{Rappaport1983} over \citetalias{Matt2015}~MB. However, a larger number of spectroscopically constrained systems is needed to draw a firm conclusion.

\section{Alternative Formation Channels}
\label{sec:alternative formation channels}

Sections~\ref{sec:GWR only} \& \ref{sec:MB} showed models where the donor descends from the WD channel. However, these arguments apply equally well to the other two AM~CVn formation channels. Here we show example model tracks from the evolved~CV (Section~\ref{subsec:evolved CV}) and He star (Section~\ref{subsec:He star}) channels, and demonstrate that the accretor temperature can place strong constraints on the extra AML (through MB) regardless of progenitor channel. 


\subsection{Evolved CV Channel}
\label{subsec:evolved CV}

As examples of AM~CVn binaries descending from the evolved~CV channel, we evolved binaries with an initially $1.1\,\msun$ ZAMS star, and an initially $0.75\,\msun$ accretor. As described in Section~\ref{sec:numerics}, the binary evolution is performed with $\mesa$ version \verb|r12778| whereas the detailed accretor evolution is handed off to version \verb|24.08.1|. We assumed that when the donor surface is H-rich ($X_{\rm s} > 0.1$), nova outbursts on the accretor expel all the transferred mass from the binary as a fast wind carrying the specific orbital angular momentum of the accretor, and once $X_{\rm s} \leqslant 0.1$, the mass transfer becomes conservative. 
At this point, the donor mass is $\approx 0.1\,\msun$, so ultimately the accretor grows to $\approx 0.8 - 0.85\,\msun$, in agreement with eclipse observations (e.g., \citetalias{vanRoestel2022}).

Once the transferred mass is nearly pure He, classical novae will not occur, so we assume fully conservative mass transfer. Varying the value of $X_{\rm s}$ produced small changes near period minimum but almost no change at larger periods.

We applied \carb until $X_{\rm s} \leqslant 0.1$, roughly near period minimum, and here we calculated $\tauconv$ following the same procedures as \cite{Van2019}, unlike in Section~\ref{sec:MB} where we followed \cite{Gossage2025}. These choices are made for comparisons with \cite{Belloni2023}.

Finally, once $X_{\rm s} \leqslant 0.1$, we evolved the binary with different MB prescriptions similarly to Section~\ref{sec:MB}. 
The goal of this exercise is to simply obtain an AM~CVn model through this channel, and constrain the MB after period minimum; whether or not AM~CVn systems can form through this channel, and whether \carb operates prior to period minimum (right before the AM~CVn phase), are beyond our scope.

Up until $X_{\rm s} \leqslant 0.1$ (near period minimum), our models are in essence similar to \cite{Belloni2023}. One major difference is that we adopted an \aesopus~low-temperature opacity table tailored for evolved~CV systems, covering a broad range of H and CNO mass fractions (see Appendix~\ref{sec:appendix opacity table}), whereas \cite{Belloni2023} adopted \cite{Freedman2008} opacities which assume a fixed H-rich composition. 
This difference in low-temperature opacities drives a shift in the parameter space required to form AM~CVn systems. Comparing with \cite{Belloni2023}, we found that there are some differences in the donor mass-radius relation near/after period minimum, which we attribute to different opacities and EOS tables being used. However, these differences do not change our conclusions.

We assume an initial accretor central temperature $\iniTwd = 10^{7}$~K, which is in the range of equilibrium core temperatures $\approx (0.4-1)\times10^{7}$~K expected for classical novae \citep[][]{Townsley2004}. Adopting a lower $\iniTwd = 5 \times 10^{6}$~K does not change our results, especially at long $\Porb$. For models shown in Appendix~\ref{sec:appendix eCV GWR}, the difference in $\Teffwd$ is $30\%$ at $\Porb=40$~min, and decreases to $\lesssim 10\%$ at $\Porb=50$~min. 


\subsubsection{Evolution with MB}
\label{subsubsec:eCV MB}

\begin{figure*}[t!]
\centering
\fig{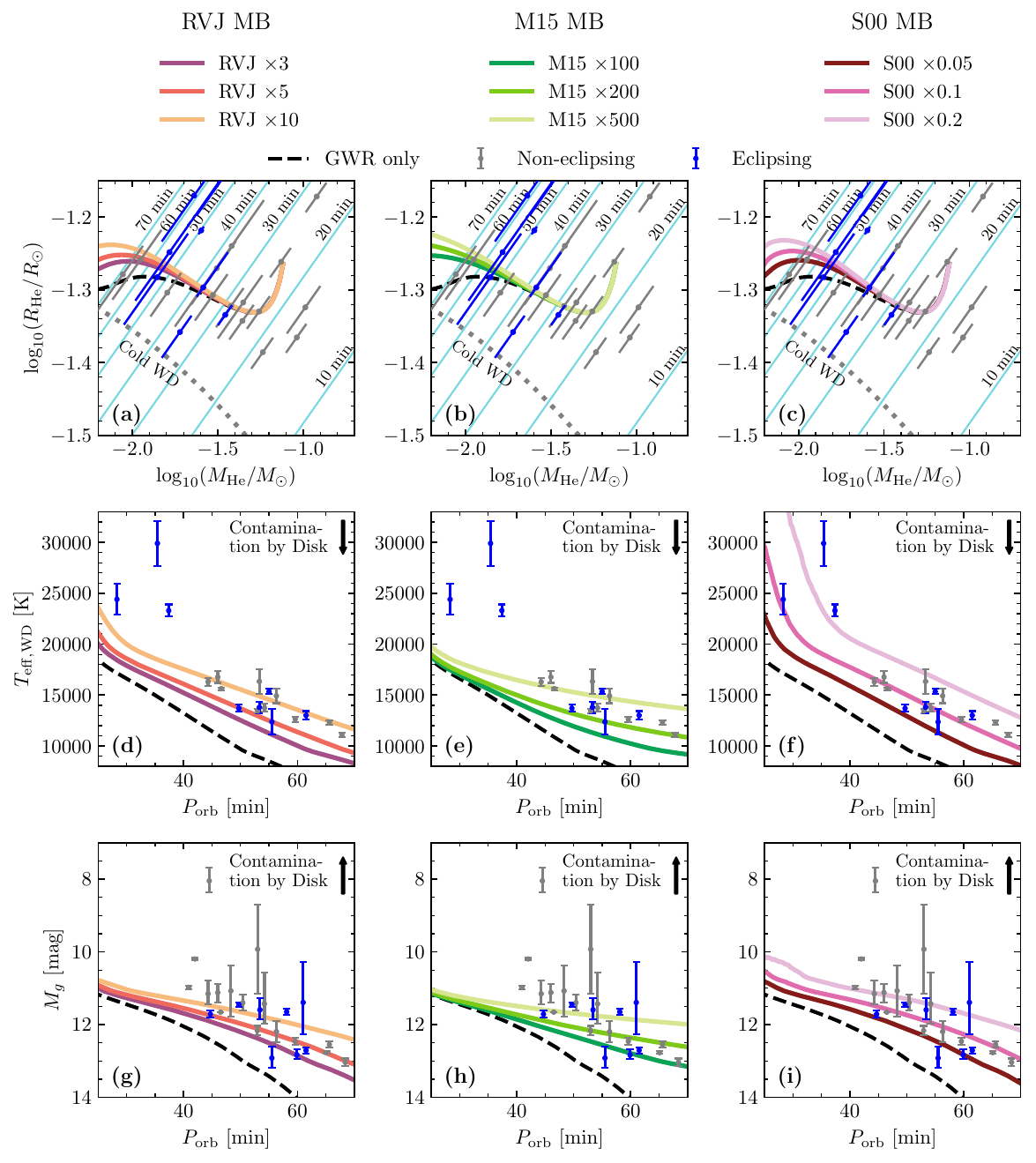}{ \textwidth }{}
\vspace{-25 pt}
\caption{ 
Same as Fig.~\ref{fig:MB_evolution2_observables}, but for the evolved CV channel. 
\label{fig:MB_evolution_eCV}}
\end{figure*}

In the evolved~CV channel, since the binary can evolve to a longer period minimum compared to the WD channel, the peak $\mdot$ is lower, leading to less efficient reheating of the accretor. 
We show this in detail in Appendix~\ref{sec:appendix eCV GWR}, for GWR-only models with different period minima. 
These models similarly show that GWR-only models underpredict $\Teffwd$ compared to observations, which can be mitigated with extra AML. 
Now the question is, how much extra AML is needed? 

We took the $\iniP = 1$~d model at $X_{\rm s} = 0.1$ near period minimum, and evolved it forward with various amounts of MB while allowing the donor to cool, similarly to Section~\ref{subsec:varying MB strength}. This model was chosen because its mass-radius relation agrees with the majority of observations. 

The top row of Figure~\ref{fig:MB_evolution_eCV} shows the donor mass-radius relation after period minimum. 
In all these models, the transition to a fully convective donor happens later at $\Porb \gtrsim 70$~min, in part due to the faster orbital evolution. 
This results in excellent agreement with observations. 
Compared to those from the WD channel in Figure~\ref{fig:MB_evolution2_observables}, the donor cooling occurs at even longer $\Porb$, because the higher surface opacity of the $1.1\,\msun$ progenitor slows down the donor cooling (see Section~\ref{subsec:donor opacity}). 

The \sills prescription yields models that cool sharply with $\Porb$, whereas those with \matt cool gradually, and those with \rvj lie in between. This is due to the the scalings of the different MB prescriptions: $\JdotSills \appropto \Porb^{-5}$, so the orbital evolution significantly slows down at long $\Porb$, allowing the donor to cool; in contrast, $\JdotRVJ$ and $\JdotMatt$ vary more shallowly as $\appropto \Porb^{-3}$ and $\Porb^{-1}$, leading to more gradual donor cooling. 

The middle and bottom rows of Figure~\ref{fig:MB_evolution_eCV} show the accretor effective temperature, and accretor $g$-band absolute magnitude, respectively. Below we discuss the results of the various MB prescriptions separately. We remind the reader that contamination by the accretion disk is stronger towards shorter $\Porb$, especially $\lesssim 50$~min, which our models do not account for (see Section~\ref{subsec:varying MB strength}). 

\bigskip

\textbf{\citetalias{Rappaport1983}~MB}: 
Models boosted by factors $\approx 5 - 10$ appear to reproduce the $\Teffwd-\Porb$ and $M_{g}-\Porb$ relations well, with the former favoring slightly higher boosting factors $\approx 10$ and the latter slightly lower boosting factors $\approx 5$. The model with boosting factor of $ 5$ appears to underpredict $M_{g}$ at $\Porb \approx 40 - 50$~min, but this is likely due to contamination by the disk.

All models appear to underpredict the $\Teffwd$ of the $\Porb = 28 - 37$~min systems derived from HST spectroscopy. 
If these systems reflect the general trend, then the evolved~CV channel requires a MB prescription that scales more steeply than $\Jdotmb\appropto \Porb^{-3}$. This is in contrast with the WD channel, where we found that boosted \rvj models can reproduce the properties of the HST-observed systems (Section~\ref{subsec:varying MB strength}). This is a consequence of the longer period minimum of the evolved CV channel: the peak $\mdot$ is lower, leading to less efficient reheating of the accretor, and so $\Teffwd$ is lower especially at short $\Porb$ where the reheating is more important compared to WD cooling. 

\bigskip

\textbf{\citetalias{Matt2015}~MB}: 
We find that \matt models boosted by $\approx 200$ can roughly reproduce the observed $\Teffwd$ and $M_{g}$ at $\Porb \gtrsim 60$~min. However, some discrepancies exist at shorter $\Porb$. For one, the models cannot reproduce the spectroscopically derived $\Teffwd$ of the $\Porb = 28 - 37$~min systems. Secondly, the models underpredict $\Teffwd$ at $\Porb\lesssim 50$~min. There we expect the disk contamination to skew the inferred $\Teffwd$ towards lower values, so the actual discrepancy may be even more severe. Both of these suggest that the evolved CV channel requires $\Jdotmb$ that scales more steeply than $\JdotMatt \appropto \Porb^{-1}$. 

\bigskip

\textbf{\citetalias{Sills2000}~MB}: 
Models scaled by $\approx 0.1$ can reproduce the $\Teffwd$ and $M_{g}$ trends shown by observations. While the model scaled by $0.1$ yields excellent agreement with the general $\Teffwd$ trend and YZ~LMi at 28~min, contamination by the accretion disk at short $\Porb$ means that the actual $\Teffwd$ is higher than shown by this model. The model with boosting factor of $0.2$ can reasonably reproduce the $\Teffwd$ of ZTF J0407-0007 and J2252-0519 at $35 - 37$~min, but overpredicts $\Teffwd$ and $M_{g}$ of the other systems. An intermediate boosting factor of, say, $0.15$, may be preferred. Overall, for the evolved~CV channel, \sills seems to reproduce many observed properties.


\subsection{He Star Channel}
\label{subsec:He star}

\begin{figure*}[t!]
\centering
\fig{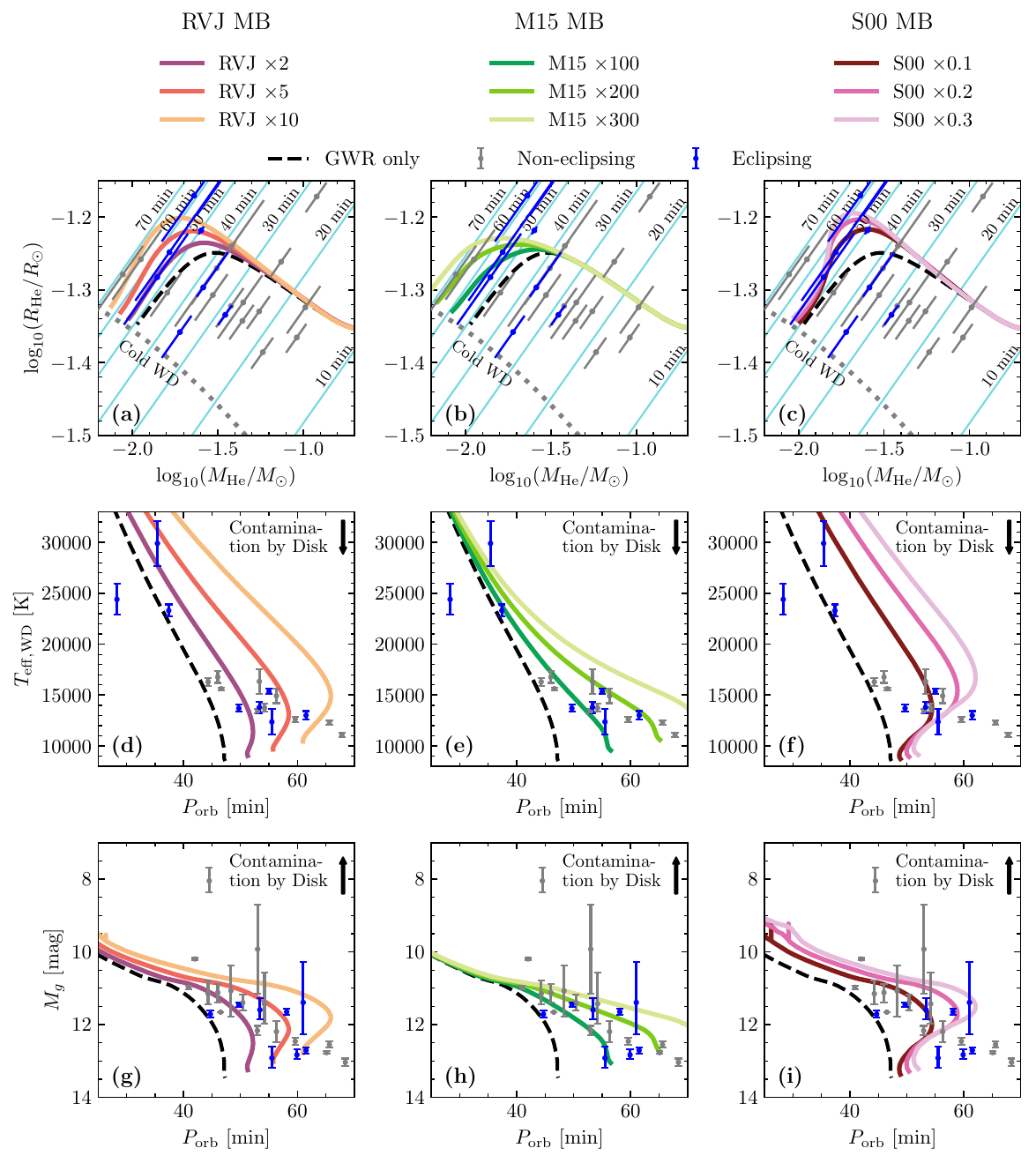}{ \textwidth }{}
\vspace{-25 pt}
\caption{ 
For a He star donor, donor mass-radius relation (top row), accretor effective temperature (middle row), and $g$-band absolute magnitude (bottom row), for various MB prescriptions scaled by different factors. Data points and other annotations same as Figure~\ref{fig:MB_evolution2_observables}. 
\label{fig:MB_evolution_Hestar_observables}}
\end{figure*}

In the He star channel, the donor is a helium-burning star that initiates 
mass transfer while still on, or shortly after departing, the He-burning sequence. As mass transfer proceeds and the donor drops below the 
He-burning mass limit ($\approx 0.32\,M_\odot$; e.g. \citealt{Savonije1986, Yungelson2008}), He burning extinguishes and the donor 
subsequently evolves as a cooling, semidegenerate object, analogous to 
the WD channel but with a distinct thermal and compositional history.

As our fiducial He star model, we adopt a $0.35\,M_\odot$ donor with an 
initial orbital period of $P_\mathrm{orb}^i = 20$ min, with a composition 
corresponding to the core of a post-CNO-burning $2.5\,M_\odot$ star --- in 
the mass range expected to ignite He under nondegenerate conditions 
\citep[e.g.,][]{Scherbak2023}. We adopted a $0.5\,\msun$ accretor such that its final mass $\approx 0.8 - 0.85\,\msun$ is consistent with observations. We obtained \aesopus~low-temperature radiative opacity tables with compositions appropriate for this He star model (see Appendix~\ref{sec:appendix opacity table}). We include MB in the model only after reaching the period minimum of about 11 minutes. 
This choice is made to isolate the post-period minimum evolution. Keeping MB on at all times reduces the duration of helium core burning, and in turn a $\approx 0.01$ dex difference in donor radius and a $\approx 1000$~K difference ($\lesssim 5\%$) in the accretor effective temperature for most of the evolution, which does not impact our conclusions. 

The accretor models for the He star channel do not consider any nuclear burning, so as to isolate the effects of accretion reheating and cooling. While a He flash is likely \citep[see][]{Piersanti2024,Piersanti2026,Rajamuthukumar2025}, evolving through it is numerically challenging. The main effect of a He flash is to expel mass and reduce the He envelope of the accretor. To estimate its effects, we re-ran our models with fully non-conservative mass transfer until the He star mass decreases to $0.25\,\msun$, after which a He flash does not occur in the accretor. We found a reduction in $\Teffwd$ by $\approx 10\%$, which does not impact our conclusions.

\subsubsection{Evolution with Varying MB Strength}

Similar to Fig \ref{fig:MB_evolution2_observables}, Fig. \ref{fig:MB_evolution_Hestar_observables} shows the donor mass-radius relation (top row), 
accretor effective temperature $T_\mathrm{eff,WD}$ (middle row), and 
Pan-STARRS1 $g$-band absolute magnitude $M_g$ (bottom row) for the He 
star channel. 
We test \citetalias{Rappaport1983}, \citetalias{Matt2015}, and \citetalias{Sills2000}~MB prescriptions scaled by various normalization 
factors, alongside a baseline model with only GWR-driven AML (black dashed 
line). The observational data and disk-contamination caveats discussed in 
Section~\ref{subsec:varying MB strength} apply equally 
here. 
In these models, the fully convective transition of the donor occurs at $\Porb\approx 45 - 60$~min, which produces a kink in the $\Teffwd$ and $M_{g}$ plots. Despite having larger thermal content, the He stars also have larger luminosities, leading to a shorter thermal timescale compared to the WD channel, which leads to a slightly earlier onset of the fully convective transition. 

For all MB prescriptions and with GWR only, the He star donor occupies a somewhat different space in the mass-radius diagram than the WD-channel donor at 
comparable $P_\mathrm{orb}$. The donor tends to be less compact as a function of its mass, consistent 
with a less degenerate, thermally younger object. 
This leads to a larger mass transfer rate at fixed period due to a larger mass ratio, compared to the WD channel. 
Despite these differences, the fundamental tension between GWR-only models and the observed accretor luminosities persists: the GWR-only model predicts an accretor that is systematically cooler and fainter than observed at $P_\mathrm{orb} \gtrsim 45$ min (see Figure~\ref{fig:MB_evolution_Hestar_observables}, panels d 
and g), as was found for the WD channel in Section~\ref{sec:GWR only}.

\bigskip

\textbf{\citetalias{Rappaport1983}~MB}: Boosting \citetalias{Rappaport1983}~MB by factors of $2-10$ accelerates the 
orbital evolution and delays the cooling of the He star donor. A boost 
factor of $\approx 2-5$ yields a satisfactory match to both the observed 
$T_\mathrm{eff,WD}$--$P_\mathrm{orb}$ relation, while a boost of $\approx 5 - 10$ matches  the $M_g$--$P_\mathrm{orb}$ 
relation across the well-observed period range 
($P_\mathrm{orb} \approx 45$--$65$ min). With these boost factors, the fully convective transition of the donor is delayed to 
$P_\mathrm{orb} \approx 50$--$65$ min, reproducing the inflated donor 
radii across most of the observed period range.

\bigskip

\textbf{\citetalias{Matt2015}~MB}: Scaling \matt by factors of $100 - 200$ produces models 
broadly consistent with observations. As in the WD channel, the 
$P_\mathrm{orb}^{-1}$ scaling in the saturated regime means the M15 torque 
contributes negligibly near the period minimum, with its relative importance growing at longer $P_\mathrm{orb}$. 
As discussed in 
Section~\ref{subsec:varying MB strength}, disk contamination at shorter $P_\mathrm{orb}$ 
likely causes an underestimate of the observed $T_\mathrm{eff,WD}$, so the 
true $T_\mathrm{eff,WD}$--$P_\mathrm{orb}$ relation is steeper than 
inferred, broadly consistent with these models.

\bigskip

\textbf{\citetalias{Sills2000}~MB}: Scaling \sills by 
factors of $0.1 - 0.3$ produces strong AML at short $P_\mathrm{orb}$ 
that prevents donor cooling out to $P_\mathrm{orb} \approx 45$ min. 
However, the torque weakens significantly at longer periods as 
$\tau_\mathrm{conv}$ rises steeply once the donor cools and becomes fully 
convective (by $P_\mathrm{orb} \approx 45$--$50$ min), causing the models 
to converge toward the GWR-only track. This limitation is more severe than 
in the WD channel, where S00 MB could at least account for systems out to 
$P_\mathrm{orb} \approx 55$--$60$ min. Exploring a broader range of He 
star masses and compositions, as well as alternative prescriptions for 
$\tau_\mathrm{conv}$, may shift these results, which should be the subject of future work examining the He star channel.

\bigskip

\textbf{Summary of the two channels}: Comparing the evolved~CV and He~star channel results to those of the WD channel 
(Section~\ref{subsec:varying MB strength}), we find that the required MB scale factors 
are similar across all prescriptions tested. This relative insensitivity to the 
formation channel suggests that the inferred extra AML may reflect a 
universal property of the AM CVn population at long $P_\mathrm{orb}$, 
rather than a feature tied to the specific progenitor history. 
Observationally distinguishing these channels 
requires abundance measurements of the accreted material 
\citep[e.g.,][]{Nelemans2010, Rodriguez2023}, and the accretor 
temperature provides a complementary, channel-agnostic constraint on 
the orbital evolution.


\section{Other Observational Implications}
\label{sec:implications}

A faster orbital evolution as proposed here, can in principle be tested by other observational means: we expect (1) a larger orbital period derivative $\Pdot$, and mass transfer rate $\mdot$, especially at longer $\Porb \gtrsim 40$~min, and (2) the period distribution to be less skewed towards longer $\Porb$. 

\subsection{$\Pdot$ and X-ray Luminosity}
\label{subsec:pdot}

\begin{figure*}
    \centering
    \includegraphics[width=\textwidth]{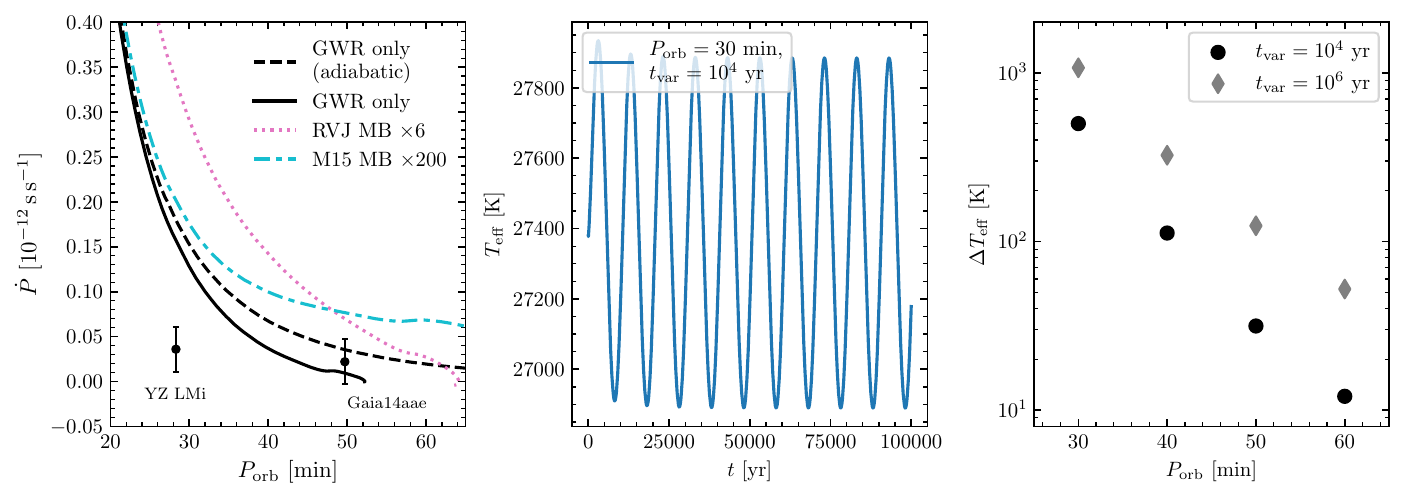}
    \caption{ 
    \textit{Left}: 
    secular period derivative $\Pdot$ of our models compared to observations. Black lines are GWR-only models, with (solid) and without cooling (dashed). Pink dotted line shows a model including \rvj boosted by 6 times, and blue dashed line shows one including \matt boosted by 200 times. 
    \textit{Middle}: $\Teffwd$ evolution of a model at $\Porb=30$~min, where we vary the accretion rate $\mdot$ sinusoidally by order unity from its secular value, on a $10^{4}$~yr timescale. 
    \textit{Right}: $\Teffwd$ variations as a function of $\Porb$, resulting from order-unity $\mdot$ variations on $10^{4}$~yr (black circle) and $10^{6}$~yr (grey diamond) timescales. 
    }
    \label{fig:Pdot_and_Teff_variations}
\end{figure*}

In principle, the faster orbital evolution as suggested by our work on $\Teffwd$ should result in a larger $\Pdot$ and $\mdot$. Compared to GWR-only tracks, the $\Pdot$ and $\mdot$ at a given $\Porb$ should be larger by $\approx10\%$ at short $\Porb\lesssim 20$~min, and up to a factor of a few at $\Porb\approx60$~min. 
A larger $\Pdot$ should be reflected in long-term eclipse timing studies \citep[e.g.,][]{Green2026}, whereas a larger $\mdot$ should be reflected in a larger accretion luminosity, potentially in X-rays \citep[e.g.,][]{Begari2023,Rodriguez2025}. 

Recently \cite{Green2026} measured the eclipse timing delay over a $\approx 10$~yr baseline for two AM~CVn binaries, YZ~LMi ($\Porb=28.3$~min) and Gaia14aae ($\Porb=49.7$~min). 
The left panel of Figure~\ref{fig:Pdot_and_Teff_variations} shows our predicted \textit{secular} $\Pdot$ compared to their  measured values. We show two GWR-only models from Section~\ref{subsec:adiabatic}, one with and one without donor cooling, and two example MB models from Section~\ref{subsec:varying MB strength} that agree well with $\Teffwd$ observations. 

All models are strongly in tension with YZ~LMi, and the ones with MB are slightly in tension with Gaia14aae. 
For YZ~LMi, the observed $\Pdot = (3.6\pm 2.5)\times 10^{-14}\,\pdotunit$ is much smaller compared to predicted, even for the GWR-only models ($\approx 2\times10^{-13}\,\pdotunit$). The \matt model results in a similar $\Pdot$ prediction as the GWR-only models, since $\JdotMatt \appropto \Porb^{-1}$ is only much stronger compared to GWR at long $\Porb$. The \rvj model, by virtue of its steeper $\JdotRVJ\appropto\Porb^{-3}$ dependence, results in a $\Pdot$ larger by $\approx 2$ times compared to the other models, and is in greater tension with YZ~LMi.

On the other hand, the three systems discovered from SRG/eROSITA \citep[][]{Sunyaev2021,Merloni2024} data by \cite{Rodriguez2025} have higher X-ray luminosities than GWR-only models, even with an optimistic radiative efficiency of $0.3$ (see their Figure~4). This led \cite{Rodriguez2025} to favor the evolved~CV models with MB by \cite{Sarkar2023_eCV}. We note, however, that this can be remedied with order-unity variations in $\mdot$ (see discussion below) in the GWR-only case with a lower radiative efficiency of $\approx 0.1$, since the X-ray luminosity scales directly with $\mdot$. Alternatively, the discrepancy can be remedied with a faster orbital evolution and hence a higher $\mdot$ regardless of progenitor channels, as we have shown in this paper. 
Our MB models, such as \rvj scaled by 6 times or \matt scaled by 200 times, are in better agreement with \cite{Rodriguez2025} assuming a radiative efficiency of $\approx 0.05 - 0.1$. 
A full comparison is difficult, due to the small number of eRosita systems at the moment and large uncertainty in the radiative efficiency. 

However, both the $\Pdot$ and X-ray luminosity measurements probe the \textit{instantaneous} $\mdot$, which may differ on short timescales from the \textit{secular} $\mdot$ predicted by our models. 
This is naturally expected, since the instantaneous $\mdot$ depends exponentially on the donor atmospheric scale height $H$, and any small changes in the stellar radius of order $H$ would result in a huge change in $\mdot$ \citep[e.g.,][]{Frank2002,King2024}. 
This deviation should occur on timescales $\sim (H/\Rhe) (\mhe/\mdot) \sim 10^{4}$~yr at $\Porb \approx 30$~min, and even longer for longer $\Porb$, much longer than any observational campaign 
\citep[e.g.,][]{Green2026}.

Alternatively, the short-term variations can be explained by changes in the stellar quadrupole moment and the stellar angular momentum distribution, known as the Applegate mechanism \citep[][]{Applegate1992}. This is argued to be the driving mechanism for the eclipse-timing variations seen in CVs \citep[e.g.,][]{Bours2016}, which exhibit $\Pdot$ that deviate from theoretical predictions sometimes by orders of magnitude and even with the wrong sign \citep[e.g.,][]{Schaefer2024}. 

Either way, in the face of short-term changes in $\mdot$, how does $\Teffwd$ deviate from the secular trend? 
Following \cite{Townsley2009} in the context of CVs, we constructed numerical experiments in which we evolved the accretor but varied the accretion rate $\mdot$ from its secular value sinusoidally by order unity, on a $t_{\rm var} = 10^{4}-10^{6}$~yr timescale. 

The middle panel of Figure~\ref{fig:Pdot_and_Teff_variations} shows an example, where we saved the \rvj model boosted by 6 times once it reaches $\Porb = 30$~min, and applied a $\mdot$ that varies sinusoidally on a $10^{4}$-yr timescale, for ten cycles. 
The average $\langle\mdot\rangle \approx 2.4 \times 10^{-10}\,\msunyr$, and given order unity changes in this $\mdot$, $\Teffwd$ deviates from its average value of 27400~K by only $\approx500$~K. 
The implied luminosity difference $\Delta L$ is given, to within factor of order unity, by the luminosity contribution of the layer at which the local thermal timescale $t_{\rm th}$ is roughly equal to $t_{\rm var}$, i.e. $\Delta L \approx L_{\rm surf} - L(t_{\rm th} = t_{\rm var})$, where $L_{\rm surf}$ is the surface luminosity of the equilibrium model \citep[][]{Townsley2009}. 

The right panel of Figure~\ref{fig:Pdot_and_Teff_variations} shows the scatter in $\Teffwd$ given our imposed short-term $\mdot$ variations. 
The base model is evolved with \rvj boosted by 6 times. 
Towards longer $\Porb$, the importance of compressional heating decreases with $\langle\mdot\rangle$, leading to a smaller $\Teffwd$ variation. For order unity variations on a $10^{4}$-yr timescale, the $\Teffwd$ variations decrease from $\approx 500$~K to just $ 10$~K. For a longer $10^{6}$-yr timescale, which is more likely for longer $\Porb$, the $\Teffwd$ variations decreases from $\approx 1000$~K at $\Porb \approx 30$~min to $ 50$~K at $\Porb \approx 60$~min, much smaller than the average value. 

We argue that $\Teffwd$ is an excellent tracer of the secular orbital evolution, with small deviations possibly arising from short-term $\mdot$ variations. With a large sample and better $\Teffwd$ measurements, we can place better constraints on the orbital evolution of the broader population. Deviations in $\Teffwd$ from the general trend can then be tied to deviations in $\Pdot$ and $\mdot$. We conjecture that YZ LMi, given its small $\Pdot$ \citep{Green2026}, may show a $\Teff$ slightly lower compared to other systems at similar $\Porb$, which future observations will test.


\subsection{Orbital Period Distribution}
\label{subsec:period distribution}

\begin{figure}
    \centering
    \includegraphics[width=0.5\textwidth]{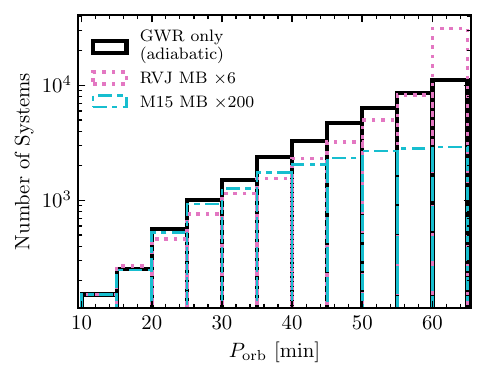}
    \caption{Steady-state period distribution of models evolved with GWR only and adiabatic evolution of the donor (black solid), with \rvj boosted by 6 times (pink dashed), and with \matt boosted by 200 times (blue dot-dashed). }
    \label{fig:period distribution}
\end{figure}

Figure~\ref{fig:period distribution} shows the steady-state orbital period distributions of various models. 
The normalization is arbitrarily chosen, such that all models have the same number of systems in the first bin. 
In all cases, the period distribution increases towards longer $\Porb$, because of the increasing evolutionary timescale. However, current detection methods are biased against these long-period systems, due to their low optical luminosity and low photometric variability \citep[see Fig.~2 of][]{Green2025}. As we explain below, future works building a complete sample corrected for observational bias\textbf{es}, can reveal the underlying orbital evolution. 

With the MB models considered here, the period distribution is less skewed towards long $\Porb$. Compared to GWR-only models, these models have a weaker dependence of $\Jdot$ with $\Porb$, and hence stronger $\Jdot$ at long $\Porb$, which means they produce fewer systems there. 
Comparing the GWR-only model (black solid line) and the \matt model boosted by 200 times (blue dot-dashed line), the number of systems in the latter can be factor of a few lower. The \rvj model boosted by 6 times also shows a faster orbital evolution, as can be seen in the lower number of systems between $\Porb \approx 30$ and 55 min, compared to the GWR-only case. However, beyond $\Porb=55$~min, cooling of the donor slows down the orbital evolution, leading to a pileup of systems at the longest period bin, as was pointed out by \cite{Deloye2007}. This may be mitigated by irradiation of the donor and a higher opacity at the donor surface (see Sections~\ref{subsec:irradiation}~\&~\ref{subsec:opacity}). The overall point still stands: evolution with the MB formalisms considered here should lead to fewer long-period systems.

\section{Conclusion}
\label{sec:conclusion}

We explored the orbital evolution of AM~CVn binaries, accounting for the stellar evolution of both the donor and \st{the }accretor WDs. 
Our main findings are as follows:
\begin{enumerate}

    \item \textit{The accretor effective temperature is an excellent tracer of orbital evolution.} The accretor is reheated at short $\Porb$ and later cools at long $\Porb \gtrsim 40$~min, yielding a relation between $\Porb$ and $\Teffwd$ (Section~\ref{subsec:general features}). In the face of short-term $\mdot$ deviations from the secular trend, $\Teffwd$ changes by less than a few percent, and thus traces the long-term orbital evolution, as opposed to short-term tracers like period derivative and X-ray luminosity (Section~\ref{subsec:pdot}). 
    
    \item \textit{GWR-only models underpredict the accretor temperature at long periods.} Eclipsing studies show that at long $\Porb \gtrsim 50$~min, the accretor dominates the luminosity of the system (e.g., \citealt{Green2018}, \citetalias{vanRoestel2022}), enabling a direct comparison with our theoretical models. Models that include GWR as the sole AML mechanism underpredict the accretor temperature  (see also \citealt{Ramsay2018}, \citetalias{vanRoestel2022}, \citealt{Macrie2024}). Cooling of the donor slows the orbital evolution and leads to a lower accretor temperature at a given $\Porb$. This can be slightly mitigated with irradiation of the donor and a higher opacity at its surface \citep[][]{Deloye2007}, but ultimately cannot reconcile the discrepancies with observations (Sections~\ref{subsec:irradiation}~\&~\ref{subsec:opacity}). 

    \item \textit{A faster orbital evolution, provided by magnetic braking of the donor, agrees better with observations. } With a faster orbital evolution, the accretor is younger (hence more luminous) at a given $\Porb$, in better agreement with observations. 

    \item \textit{The required extra torque becomes increasingly important relative to GWR at long $\Porb$, and grows from a few to tens of percent at short periods, to a few times greater than GWR at $\Porb \approx 60$~min}. We tested various MB prescriptions, finding that none can reproduce the observed accretor luminosity (Section~\ref{subsec:original MB}). We then scaled three popular MB prescriptions (\citetalias{Rappaport1983}, \citetalias{Matt2015}, \& \citetalias{Sills2000}) . We found that \rvj ($\Jdot \appropto \Porb^{-3}$) boosted by $\approx 3 - 10$ times, or \matt ($\Jdot \appropto \Porb^{-1}$) boosted by $\approx 80 - 250$ times, can reproduce observations of the accretor luminosity (Section~\ref{subsec:varying MB strength}). Despite the different scalings, these prescriptions yield a torque $\Jdotmb$ a few times that of GWR (but, more robustly, no more than $\approx 10 $ times) at long $\Porb \approx 60$~min. 

    \item \textit{Other progenitor channels}. While our fiducial models are for the WD channel, we considered models from the He star and evolved CV channels (Section~\ref{sec:alternative formation channels}). We found that the required extra torque is roughly the same across all three progenitor channels. 
     
\end{enumerate}

This work shows that the accretor temperature is a powerful tool to constrain the orbital evolution of AM~CVn binaries, and joins a growing number of studies suggesting that MB is boosted in close binaries \citep[e.g.,][]{Blomberg2024,Belloni2024}. 
Extrapolating our results to the shortest period $\Porb \lesssim 10$~min, we expect a $\Jdot$ that is a few \% up to 50\% greater than GWR-only, comparable to the effects of tides \citep[e.g.,][]{Piro2019}. This will impact the chirp signal detected by LISA, and should be considered in future studies. 

Our comparison to observations is hampered by the lack of spectroscopic models for AM~CVn accretors, necessitating the use of blackbody or DB WD atmosphere assumptions \citep[e.g.,][]{Macrie2024}. Future spectroscopically-derived accretor temperatures \citep[e.g.,][]{Yu2026,Yu2026b} will yield much better constraints on the accretor temperature, and hence the orbital evolution. Our models are also limited by uncertainties in the opacity of the donor, which future models will improve upon (see Appendix~\ref{sec:appendix opacity table}). 
The extra MB torque we inferred is an empirical result, but should be considered in future, physically motivated theories of MB. 

Extra AML in AM~CVn binaries can in principle originate from MB by the accretor as well \citep[][]{Farmer2010}. Several systems are thought to host a magnetic accretor, including Gaia14aae which has the best constrained donor mass and radius \citep[][]{Maccarone2024,ElBadry2026}. We defer to future studies to investigate the impact of accretor MB on the orbital evolution.

\section*{Acknowledgments}

We thank the referee for their constructive suggestions that have greatly improved our manuscript. 
We are grateful to Lars Bildsten, Kareem El-Badry, Kevin Burdge, Matthew Green, Tony Piro, Jan van Roestel, and Weitian Yu, for stimulating conversations about interpreting observations of AM CVn binaries. 
We thank Simon Blouin, Shigenobu Hirose, and Linhao Ma for helpful conversations regarding opacities. 
We appreciate conversations with Seth Gossage on magnetic braking, and Diogo Belloni and Mattias Schreiber on the evolved CV channel. 
We appreciate the continuous effort by the {\mesa} developing team in maintaining the code and providing new features, and by the {\aesopus} team in providing open-access opacity calculations. 
Discussions for this work were supported, in part, by the Gordon and Betty Moore Foundation through grant GBMF5076. 
This publication makes use of VOSA, developed under the Spanish Virtual Observatory project supported from the Spanish MICINN through grant AyA2011-24052. PS is grateful for support from the Gordon and Betty Moore Foundation through Grant GBMF5076. 

\software{
\texttt{MESA} \citep[r12778, 24.08.1;][]{MESAI,MESAII,MESAIII,MESAIV,MESAV,MESAVI}, 
\texttt{py\_mesa\_reader} \citep{bill_wolf_2017_826958},
\texttt{ipython/jupyter} \citep{perez_2007_aa,kluyver_2016_aa},
\texttt{matplotlib} \citep{hunter_2007_aa},
\texttt{NumPy} \citep{numpy2020}, 
\texttt{SciPy} \citep{scipy2020}, 
\texttt{Astropy} \citep{astropy2013,astropy2018,astropy2022},
and 
\texttt{Python} from \href{https://www.python.org}{python.org}
}



\appendix

\twocolumngrid

\begin{figure*}[t!]
\centering
\fig{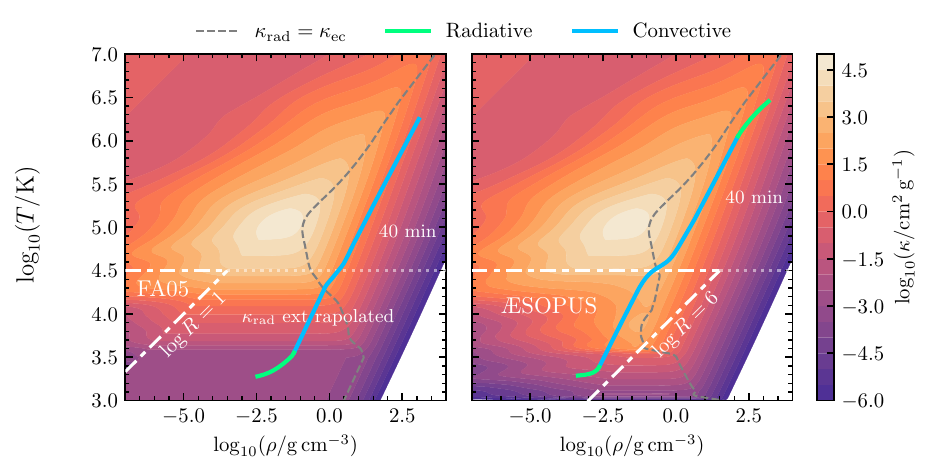}{ \textwidth }{}
\vspace{-25 pt}
\caption{ 
Opacity maps retrieved from $\mesa$ version r12778 using \citealt{Ferguson2005} (FA05; left) and \aesopus\,(right), respectively, as the low-temperature radiative components. Solid white dash-dotted lines show the regimes of validity of these tables, beyond which an extrapolation is done (below the faded white dotted lines). Grey dashed lines show where the radiative opacity $\kappa_{\rm rad}$ is equal to the electron conduction opacity $\kappa_{\rm ec}$. For illustrative purposes, we show temperature-density profiles of donors at $\Porb=40$~min, evolved using these tables. Green shows radiative zones and blue shows convective zones. 
\label{fig:kap_comparison}}
\end{figure*}

\section{\aesopus~opacity table}
\label{sec:appendix opacity table}

\textbf{Fiducial}: 
In Figure~\ref{fig:kap_comparison} we show the opacity maps retrieved from $\mesa$ with \citealt{Ferguson2005}~(FA05) and \aesopus~as the low-temperature radiative opacity component. 
The former shows the default $\mesa$ settings as adopted in \citetalias{Wong2021}, and the latter is used in this work. 
We adopted $X=0, Z = 0.0142$ with \cite{Asplund2009} abundances. For the \aesopus~tables, we further modified C, N, O, F, Ne and Na to account for the CNO and NeNa cycles on the MS, with a $1.5\,\msun$ $\mesa$ model as our reference. 
The corresponding enhancement factors $f_{\rm Z} \equiv \log_{10}(X_{\rm Z}/X_{\rm Z,\odot})$, where $X_{\rm Z}$ is the mass fraction of a particular element, and $X_{\rm Z, \odot}$ is that of the Sun, are $f_{\rm C} = -1.39$, $f_{\rm N} = 1.06$, $f_{\rm O} = -1.09$, $f_{\rm F} = -3.17$, $f_{\rm Ne} = -0.034$, and $f_{\rm Na} = 0.64$. 
Note that the \aesopus~opacities here do not account for condensation which occurs at temperatures $\lesssim 1500$~K 
\citep[e.g.,][]{Burrows1997,Allard2001,Baraffe2003,Marigo2024_condense}. This adds to the opacity uncertainties as discussed in Section~\ref{subsec:opacity}. 

The opacity tables are valid for $\log_{10}(T/\rm{K}) \leqslant 4.5$ and up to a certain $\log_{10} R = \log_{10}\rho - 3 \log_{10} T + 18 $. The FA05 opacity tables only go up to $\log_{10} R = 1$, whereas \aesopus~extends to higher densities at $\log_{10} R = 6$. At higher $\log_{10} R$, $\mesa$ takes the table value at $\log_{10}R = 1$ and the same $\log_{10} T$, which explains why the contours are flat with $\log_{10} T$ below the white dotted lines in Figure~\ref{fig:kap_comparison}. This trend discontinues at even higher densities, to the right of the grey dashed lines, where electron conduction dominates the opacity. 
Additionally, in the extrapolated regime of the FA05 opacities, for $\log_{10}(T/\rm{K}) \leqslant 3.5$ the radiative opacity is fixed by its value at $\log_{10} R = 1$ and $\log_{10}(T/\rm{K}) = 3.5$ (specifically for $\mesa$ version \verb|r12778|). 

Figure~\ref{fig:kap_comparison} shows temperature-density profiles of donors evolved with the corresponding opacity tables. All other aspects of the binary runs are identical: the donors initially have $\iniSc = 3.7\,\kB\NA$, and we allow irradiation of the donor. The profiles are taken at $\Porb = 40$~min, when the donors have experienced some amount of cooling. 
The profiles show clearly that the donor is well within the extrapolated regime for FA05 ($\log_{10} R > 1$), but still within coverage by the \aesopus~opacity tables. 
The different opacities change the location of the radiative-convective boundary, as well as the amount of cooling the donors experience. 
In general, the opacities increase with density, and due to the lack of density dependence, the extrapolated FA05 opacities are lower. In turn, the donor experiences more cooling with the extrapolated FA05 opacities. 
This is why the donor with the extrapolated FA05 opacities is already fully convective and has lower entropy. 
The thermal evolution of the donor, and in turn the orbital evolution, are affected as a result. 
This comparison shows clearly why we chose to adopt \aesopus~opacity tables, which have better coverage, for this work. 

\begin{figure}
\centering
\fig{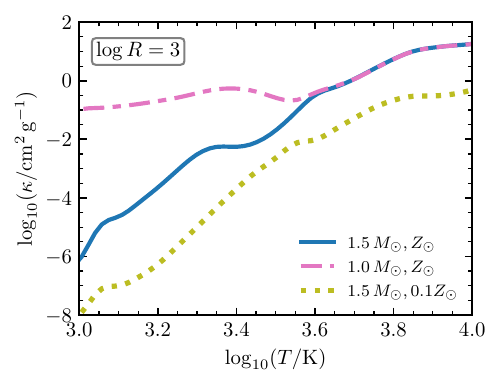}{ 0.5 \textwidth }{}
\vspace{-25 pt}
\caption{ 
Low-temperature opacities taken at $\log R = 3$ with different underlying abundance patterns and metallicities. 
Blue solid line shows the default used in this work, for a $1.5\,\msun$, solar-metallicity ($Z = 0.0142$) post-H burning star. Pink dot-dashed line shows that of a $1\,\msun$, solar-metallicity star. Yellow dotted line shows that of a $1.5\,\msun$, one-tenth solar metallicity star. 
\label{fig:opacity_mass}}
\end{figure}

\textbf{Changing progenitor mass and metallicity}: 
Figure~\ref{fig:opacity_mass} compares the low-temperature opacities taken at $\log R = 3$, for the three different abundance patterns discussed in Section~\ref{subsec:donor opacity}. 
At fixed metallicity, the opacities only differ significantly below $\approx 3000$~K. 
This is not surprising given that molecular opacities, which depend sensitively on the abundance pattern, dominate there. Above this temperature, the opacity is dominated by atomic opacities of He and metals. 
Of note are alkali metals like Na and K, which have significant line opacities \citep[e.g.,][]{Burrows2000} and furthermore contribute free electrons to the He-minus free-free opacity \citep[e.g.,][]{Blouin2018}. 
Because the alkali metals dominate at cooler temperatures until molecular opacities kick in, it is important that we included an enhancement of Na by $\approx 0.6$~dex relative to solar (in mass fraction), due to the NeNa cycle on the MS, as has been observed in evolved~CV systems \citep[][]{Yamaguchi2023}. 

At solar metallicity, the opacity at low-temperatures $\lesssim 3000$~K is a lot higher with a $1\,\msun$ progenitor, compared to that of a $1.3$ or $1.5\,\msun$ progenitor. 
This is due to their different O abundances, and hence C/O ratios. 
Across the mass range we tested, the CN~cycle during the MS leads to a depletion of C at similar levels, by almost an order of magnitude. 
For the $1\,\msun$ progenitor, the NO~cycle only sets in near the end of the MS, with limited changes to the O abundances. 
This leads to a C/O ratio much lower than unity in the $1\,\msun$ abundance pattern \citep[e.g.,][]{Toloza2023}, and the excess O goes on to form O-bearing molecules that contribute to the opacities. 
By recomputing the opacities in the absence of individual elements, we find that the O-bearing molecules of Mg, Si, S and Ca contribute significantly to the opacities, for the $1\,\msun$ abundance pattern. 
In contrast, for stars with ZAMS mass $\gtrsim 1.3\,\msun$, in addition to the CN~cycle, the NO~cycle also operates on the MS and similarly depletes O by almost an order of magnitude. Most of the O is locked in CO and SiO, and other O-bearing molecules are depleted, leading to a lower opacity \citep[e.g.,][]{Marigo2009}. 

The opacity decreases with metallicity, since the He-minus free-free opacity depends on the free electron number density coming from the metals \citep[e.g.,][]{John1994}, and the atomic metal opacities and molecular opacities all depend on the metal abundance. However, at $0.1\,Z_{\odot}$, the variation with ZAMS mass is less significant. For stars that only undergo the CN~cycle on the MS, e.g. a $1\,\msun$ star, the NO~cycle sets in near the end of the MS and modifies the O abundances as well, leading to similar central CNO abundances across different masses. 

\textbf{He star:} 
We obtained \aesopus~opacities appropriate for the composition of our $0.35\,\msun$ He star with an initial orbital period of 20~min. 
The He star has an initial composition corresponding to the core of a post-CNO burning $2.5\,\msun$ star, which is in the mass range expected to ignite He under non-degenerate conditions \citep[e.g.,][]{Scherbak2023}. 
Since mass transfer quickly brings the He star below the He-burning limit ($\approx 0.32\,\msun$), the core of the He star is modified by some triple-alpha burning, but little alpha-capture onto $^{16}\rm{O}$. As a result, C is enhanced by a factor $\approx 100$ compared to post-H burning (in mass fraction), while O is only moderately enhanced by $\approx 70\%$. 
These correspond to mass fractions of $\approx 0.01$ and $5 \times 10^{-4}$, respectively. 
Therefore, we constructed \aesopus~tables with C enhancement factors $f_{\rm C}$ ranging from $-1.4$ to $0.6$, and C/O ratio enhancement factors $f_{\rm CO}$ ranging from $-0.1$ to $1.7$. 
The limited He burning also leads to little $^{14}\rm{N}$ being consumed into $^{18}\rm{O}$. We therefore specified a fixed N enhancement factor of $f_{\rm N} = 1.065$. Similarly to the He WD models, we specified $f_{\rm F} = -2$, $f_{\rm Ne} = -0.033$, and $f_{\rm Na} = 0.609$. 

\textbf{Evolved CV:} 
An evolved CV exhibits various levels of H depletion and CNO-process enhancement throughout its evolution. We therefore constructed \aesopus~opacity tables covering the range of abundances expected. To reduce the dimensionality, we specified C, N, O, Na, Ne abundances that correlate with the H mass fraction from $0$ to $0.8$, based the the surface abundances of our $\mesa$ models.


\section{Magnetic Braking Prescriptions}
\label{sec:appendix MB}

\subsection{Theoretical Expectations}
\label{subsec:theoretical expectations}

Here we lay some groundwork for how we expect the various torques to evolve with $\Porb$. 
To first order, all these prescriptions scale with $\Porb$ as is written explicitly ($\propto \Porb^{-3}$ or $\Porb^{-1}$), and for \citetalias{Van2019} and \citetalias{Sills2000}~MB, with the convective turnover timescale, $\tauconv$. 

Since the torques scale with $\mhe$ and $\Rhe$, there is an additional, implicit scaling with $\Porb$. 
Using the Roche-lobe relation and Kepler's law, $\Rhe \propto \mhe^{1/3} a \propto \mhe^{1/3} \Porb^{2/3} $. With the mass-radius relation $\Rhe \propto \mhe^{\xi}$, we get $\mhe \propto \Porb^{2/(3 \xi - 1)} $ and $\Rhe \propto \Porb^{2\xi/(3\xi - 1)} $. Taking $\xi \approx -0.2$ as a typical value, $\mhe \appropto \Porb^{-1.25} $ and $\Rhe \appropto \Porb^{0.25}$. 
For reference, the torque due to GWR is $\Jdotgr \propto \Porb^{-7/3} (\mwd \mhe)^{2} / ( \mwd + \mhe )^{2/3} \appropto \Porb^{-7/3} \mhe^{2} \appropto \Porb^{-5} $.

The scaling of $\tauconv$ with $\Porb$ can be estimated as follows. The pressure scale height near the base of the convective zone is given by $H \sim \kB T_{\rm conv}/(\mu m_{\rm p} g)$, where $T_{\rm conv}$ is the temperature near the base, and $g = G \mhe / \Rhe^{2}$ is the surface gravity. In our models, $T_{\rm conv}$ remains roughly constant with $\Porb$, so variations in $H$ are mainly driven by that of the surface gravity, $H \propto \Rhe^{2}/\mhe$. For $\xi = -0.2$, $H \appropto \Porb^{1.75} $: as the donor loses mass and expands, its surface gravity decreases, leading to a larger pressure scale height. We can estimate the convective velocity as $v_{\rm conv} \sim (L_{\rm conv}/\rho)^{1/3}$. As the donor cools, $L_{\rm conv}$ gradually decreases, whereas the density near the base of the convection zone increases. Both contribute to a decreasing $v_{\rm conv}$. 
We thus expect that $\tauconv$ scales somewhat steeper than $\tauconv \appropto \Porb^{2} $. Indeed, for most models we found that $\tauconv \appropto \Porb^{2.5-3}$.


\subsection{The MB Prescriptions}

Here we show the MB scaling laws with calibration constants as chosen in \cite{ElBadry2022}, but rewrite them in terms of physical parameters appropriate for an AM~CVn binary at $\Porb = 60$~min, where a typical donor mass is $\mhe \approx 0.01\,\msun$ and radius $\Rhe \approx 0.05\,\rsun$. We implicitly assume that tides efficiently synchronize the donor rotation to the orbit, i.e., $\Prot = \Porb$. For a $0.8\,\msun$ accretor, the Roche-lobe-filling condition at this period yields a binary separation $a \approx 0.46\,\rsun$, orbital angular momentum $\Jorb \approx 3.5 \times 10^{49}\,\Jdotunit$, and AML by GWR $\Jdotgr \approx - 1.3 \times 10^{32} \,\Jdotunit $ \citep[][]{Peters1964,Landau1975}. The corresponding timescale for orbital change driven purely by GWR is $ \sim \Jorb / | \Jdotgr | \approx 9 \times 10^{9}$~yrs.


\subsubsection{RVJ MB}
\label{subsubsec:RVJ}

The MB formulation proposed by \citealt{Rappaport1983}, hereafter \citetalias{Rappaport1983}, is based on the Skumanich law and \cite{Verbunt1981}. It shows a steep explicit scaling with $\Porb$ as $\Jdotmb \propto \Porb^{-3}$, and with stellar radius as $R^{\gamma}$, where we take $\gamma = 4$ as is widely used in literature. The \citetalias{Rappaport1983} MB prescription has been widely used in binary evolution studies \citep[e.g.,][]{Podsiadlowski2002,Knigge2011,Istrate2016}, and is the default MB prescription in \mesa. However, rapidly rotating single and binary stars appear to spin down more slowly than predicted by \citetalias{Rappaport1983}~MB \citep[see, e.g.,][for recent studies]{ElBadry2022,Gossage2023,Sun2024}. 

In \citetalias{Rappaport1983}~MB, the torque is given by 
\begin{eqnarray}
    \JdotRVJ
    &&\approx 
    - 5.9 \times 10^{31} \left( \frac{\Porb}{60\,\rm{min}} \right)^{-3} \nonumber\\
    &&\times \left( \frac{\mhe}{0.01\,\msun} \right) \left( \frac{\Rhe}{0.05\,\rsun} \right)^{4} \, \Jdotunit,
\end{eqnarray}
roughly a factor of a few weaker than AML driven by GWR. We thus expect that the orbital evolution is only slightly faster than that driven by GWR~AML. Note that, unlike the original \citetalias{Rappaport1983} MB, we do not turn off MB when the donor becomes fully convective. In Section~\ref{subsec:original MB}, we show that the original \citetalias{Rappaport1983}~MB results in an insufficient torque to explain observational data, and allowing MB to turn off results in a stronger tension.


\subsubsection{S00 MB}
\label{subsubsec:S00}

The prescription by \citealt{Sills2000}, hereafter \citetalias{Sills2000}, shows a scaling $\propto \Porb^{-3}$ in agreement with the Skumanich law, above a critical period corresponding to a Rossby number $\Rocrit \approx 0.1 $, and more weakly as $\propto \Porb^{-1}$, in the ``saturated'' regime below the critical period. This scaling was shown to reproduce the orbital period distribution of detached low-mass MS binaries \citep[][]{ElBadry2022}. 

Here we adopt $\Pcrit = 0.1 \Psol  (\tauconv/\tauconvsol)$, where $\tauconvsol = 12.9$~d is the solar convective turnover time, and $\Psol = 28$~d is the solar rotation period. 
In our models, $\tauconv$ which is calculated self-consistently, increases from $\sim 1$~d at period minimum to $\sim 10^{3}$~d at long $\Porb$, yielding $\Pcrit \sim 10 - 10^{4}$~d~$\gg \Porb$. Therefore, MB in AM~CVn binaries is always in the saturated regime, where the torque is given by
\begin{eqnarray}
    \JdotSills 
    &&\approx 
    - 1.2 \times 10^{34} \left( \frac{\Porb}{60\,\rm{min}} \right)^{-1} \left( \frac{ \tauconv }{ 100 \, \rm{d} } \right)^{-2} \nonumber\\
    &&\times \left( \frac{\mhe}{0.01\,\msun} \right)^{-1/2} \left( \frac{\Rhe}{0.05\,\rsun} \right)^{1/2} \,\Jdotunit .
\end{eqnarray}
This is about 2 orders-of-magnitude stronger than $\Jdotgr$. The binary will evolve on a timescale $\sim \Jorb / | \JdotSills | \approx 9 \times 10^{7} $~yrs, starkly shorter than that driven by GWR. 


\subsubsection{M15 MB}
\label{subsubsec:M15}

\cite{Matt2015}, hereafter \citetalias{Matt2015}, introduced a similar ``saturated'' MB prescription with the $\Porb^{-1}$ scaling in the rapidly rotating regime. 
The torque in the saturated regime is given by
\begin{eqnarray}
    \JdotMatt
    &&= 
    - 3.9 \times 10^{30}  \left( \frac{\Porb}{60\,\rm{min}} \right)^{-1} 
    \left( \frac{\mhe}{0.01\,\msun} \right)^{1/2} \nonumber\\
    && \times \left( \frac{\Rhe}{0.05\,\rsun} \right)^{3.1} \,\Jdotunit
    ,
\end{eqnarray}
which, unlike \citetalias{Sills2000}, is independent of the convective turnover timescale. Since $\JdotMatt$ is two orders-of-magnitude weaker than $\Jdotgr$, it will not make any appreciable difference in the orbital evolution. Therefore, in Section~\ref{subsec:original MB}, we do not present models for \matt, and we return to these models in Section~\ref{subsec:varying MB strength}. 


\subsubsection{CARB MB}
\label{subsubsec:CARB}

\cite{Van2019a,Van2019} proposed the Convection And Magnetic Boosted (hereafter \citetalias{Van2019}) MB model. Unlike previous empirical prescriptions, \citetalias{Van2019}~MB is based on physical principles. They accounted for strengthening of magnetic fields with increasing Rossby number based on dynamo theory \citep[e.g.,][]{Ivanova2006}, and magnetocentrifugal effects on the \alfv radius based on magnetohydrodynamical simulations \citep[][]{Reville2015}, resulting in a torque\footnote{As noted in \cite{Soethe2021}, there is a typo in the torque formula given by \cite{Van2019}.}
\begin{eqnarray}
    \JdotCARB 
    &&= - \frac{2}{3} \mdotw^{-1/3} \Rhe^{14/3} \left( v_{\rm esc,He}^{2} + \frac{8 \pi^{2} \Rhe^{2}}{ \Porb^{2} K^{2}_{2} } \right)^{-2/3} \nonumber\\ 
    &&\times \left( \frac{2 \pi }{ P_{\rm rot,\odot} } \right) \left( \frac{ \Porb }{ P_{\rm rot, \odot} } \right)^{-11/3} \left( \frac{ \tauconv }{ \tauconvsol } \right)^{8/3} , 
\end{eqnarray}
where $\mdotw$ is the donor wind mass loss rate, $v_{\rm esc,He}$ is the donor escape velocity, $K_{2} = 0.07$ is a calibrated constant. Following \cite{Van2019} and \cite{Belloni2023}, we use the Reimers wind law \citep{Reimers1975} with a scaling factor of 1, yielding $\mdotw \sim 10^{-17}\,\msunyr$ in our simulations. However, the actual wind mass loss rate may be substantially different due to the binary environment. 

Since the donor is filling its Roche lobe, $\Rhe / a \approx 0.462[\mhe / (\mhe + \mwd)]^{1/3}$. The term $8 \pi^{2} \Rhe^{2} / (\Porb K_{2})^{2}$ can then be written as $(0.462^{3}/K_{2}^{2}) v_{\rm esc,He}^{2} \approx 20 v_{\rm esc, He}^{2} $. As the name suggests, the torque is rotationally boosted. Moreover, as we show in Section~\ref{subsec:original MB}, the rapid orbital evolution due to \citetalias{Van2019}~MB leads to almost no thermal evolution of the donor. The convective turnover time therefore remains short, $\tauconv \sim 10$~d. Putting all these together, we estimate
\begin{eqnarray}
    \JdotCARB 
    &&\approx - 1.3 \times 10^{35} \left( \frac{\Porb}{60\,\rm{min}} \right)^{-11/3} \left( \frac{ \tauconv }{ 10\,\rm{d} } \right)^{8/3} \nonumber\\ 
    &&\times \left( \frac{ \mhe }{ 0.01\,\rsun } \right)^{-2/3} \left( \frac{ \Rhe }{ 0.05\,\rsun } \right)^{16/3} \nonumber\\
    &&\times \left(\frac{\mdotw}{10^{-17}\,\msunyr}\right)^{-1/3} 
     \,\Jdotunit . 
\end{eqnarray}
This is almost three orders-of-magnitude larger than $\Jdotgr$, resulting in rapid orbital evolution in $\sim \Jorb / |\JdotCARB| \approx 9 \times 10^{6}$~yrs. 


\section{Evolved CV Evolution with Only GWR}
\label{sec:appendix eCV GWR}

Here we present evolved~CV models where the donor evolves with only GWR near/after period minimum. 
These models are very similar in character as those in Section~\ref{subsec:general features}, except that because the period minima are longer, details of the accretor reheating are different. 
We include the adiabatic $\iniSc=3.7\,\kB\NA$ model from Section~\ref{subsec:general features} as a representative for the WD donor channel, and refer to it as the WD model. 

\begin{figure*}[t!]
\centering
\fig{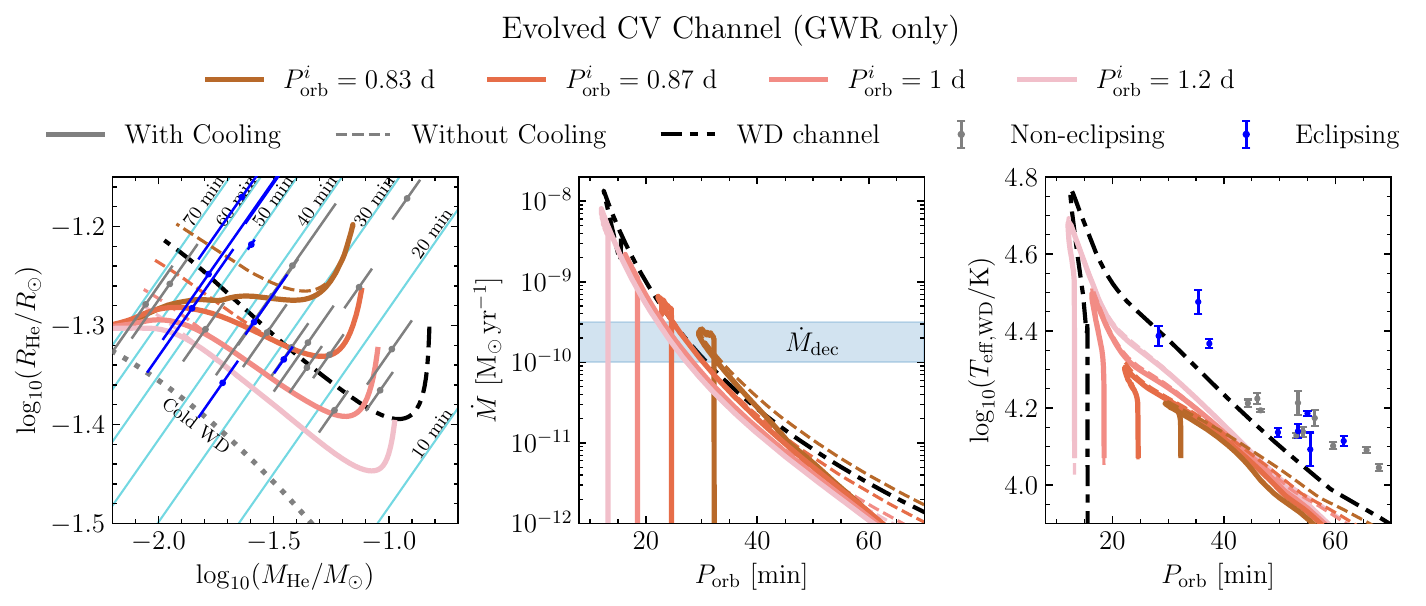}{ \textwidth }{}
\vspace{-25 pt}
\caption{ 
Similar to Figure~\ref{fig:MB_evolution}, but for evolved~CV models with GWR as the sole AML mechanism. Solid curves are evolved with cooling, and dashed curves without (i.e., adiabatically). 
The colored solid curves show evolved~CV models with initial orbital periods from $0.83-1.2$~d. The black dashed curve shows an example WD model ($\iniSc=3.7\,\kB\NA$ from Figure~\ref{fig:MR_and_Mdot_GW}). 
From left to right, the three panels show the donor mass-radius relation, mass transfer rate, and the accretor effective temperature. 
Data and other annotations same as Fig.~\ref{fig:MR_and_Mdot_GW} \& \ref{fig:Teff_and_Mg_GW}. 
\label{fig:eCV_GWR_ad}}
\end{figure*}

The left panel of Figure~\ref{fig:eCV_GWR_ad} shows the donor mass-radius relation, for models with different initial orbital periods, from $0.87$ to $1$~d. A shorter initial orbital period produces a less evolved donor. Correspondingly, the donor central entropy near period minimum is higher, and the donor has a larger radius. Consequently, the binary reaches a longer period minimum. 

The qualitative understanding of these models in terms of donor entropy, follows naturally from our discussion in Section~\ref{subsec:general features}, with minor differences stemming from the detailed thermal history and hence thermal profile of the donor. For example, the evolved~CV models that best fit the observed donor mass-radius relation ($\iniP = 0.93 - 1$~d) have central specific entropy $\Sc \approx 3.5 - 3.6 \,\kB\NA$, whereas the corresponding WD model from Section~\ref{subsubsec:GWR donor evolution} has $\Sc \approx 3.7\,\kB\NA$. However, regardless of progenitor channel, it remains true that AM~CVn donors need to have high entropy in order to reproduce the observed inflated radii. 

Once the system evolves to long $\Porb \approx 50 - 60$~min, those with cooling (solid lines) deviate from those without (i.e. the adiabatic models; dashed lines), and contract towards the fully degenerate mass-radius relation. This is because the donor thermal timescale becomes comparable to or shorter than its mass transfer timescale. These evolved~CV models cool at a longer $\Porb$ than the WD channel models (see Fig~\ref{fig:MR_and_Mdot_GW}), because they have higher opacities at the surface. As we showed in Section~\ref{subsec:donor opacity}, a lower-mass donor progenitor (e.g., $1.1\,\msun$ as we adopted for the evolved~CV channel) yields a higher opacity compared to one with higher mass (e.g., $1.5\,\msun$ as we adopted for the WD channel). 

The middle panel of Figure~\ref{fig:eCV_GWR_ad} shows the mass transfer rates. An evolved~CV system with a longer $\iniP$ (less nuclearly evolved) reaches a longer period minimum, and so produces a lower peak $\mdot$. Models that reproduce the radii of observed systems, like that with $\iniP = 1$~d, show peak $\mdot \approx 10^{-9}\,\msunyr$, almost an order of magnitude lower than the WD model. Models with more inflated radii, show even lower peak $\mdot \approx 10^{-10}\,\msunyr$. This affects the degree to which the accretor is reheated. 

The right panel of Figure~\ref{fig:eCV_GWR_ad} shows the accretor effective temperature $\Teffwd$. As expected, models with less evolved donors are reheated to a lower $\Teffwd$, due to the lower peak $\mdot$. 
However, these donors are higher entropy, and their adiabatic mass-radius relation leads to a faster orbital evolution. 
As a result, out to long $\Porb \approx 50-60$~min, the evolved~CV models show very similar $\Teffwd$ despite the different amounts of initial reheating. 
Furthermore, allowing the donor to cool (solid) yields cooler $\Teffwd$ than without (dashed), due to the former's slower orbital evolution. 

Compared to the WD model, the evolved~CV models show less core reheating. For the evolved~CV models, the accretor central temperature increases by less than $0.2$~dex, where those with less evolved donors increase even less. This is in contrast to the WD~model, where the accretor central temperature increases by $0.4$~dex. In the end, all evolved~CV models remain somewhat cooler than the WD model across all orbital periods. As with Section~\ref{subsubsec:GWR comparison to observations}, these models are cooler compared to observations, and again suggest a faster orbital evolution. 


\bibliography{main,mesa,software}{}
\bibliographystyle{aasjournalv7}

\end{document}